\PassOptionsToPackage{table}{xcolor}

\documentclass{siamart220329}

\usepackage{lipsum}
\usepackage{amsfonts}
\usepackage{graphicx}
\usepackage{epstopdf}
\ifpdf
  \DeclareGraphicsExtensions{.eps,.pdf,.png,.jpg}
\else
  \DeclareGraphicsExtensions{.eps}
\fi

\newsiamremark{remark}{Remark}
\newsiamremark{hypothesis}{Hypothesis}
\crefname{hypothesis}{Hypothesis}{Hypotheses}
\newsiamthm{claim}{Claim}

\usepackage{amsmath,amsfonts,mathtools}

\DeclareMathOperator*{\argmax}{\arg\!\max}
\usepackage{bm}

\usepackage{algpseudocode}

\Crefname{ALC@unique}{Line}{Lines}

\usepackage{siunitx}
\usepackage{multirow}
\usepackage{makecell}
\usepackage{array}

\usepackage{textcomp}

\usepackage{booktabs}

\usepackage{xcolor}

\definecolor{ZachColor}{rgb}{0.9, 0.5, 0}
\definecolor{TeseoColor}{rgb}{0.15, 0.68, 0.38}
\definecolor{DanieleColor}{rgb}{1, 0, 1}
\definecolor{DavidColor}{rgb}{.13, .67, .8}
\definecolor{MarcoColor}{rgb}{1, 0, 0.5}
\definecolor{BolunColor}{HTML}{6DC70F}
\definecolor{Violet}{rgb}{.6, .35, .7}
\definecolor{Pink}{rgb}{1, 0, 1}
\definecolor{Yellow}{rgb}{1, 0.65, 0}
\definecolor{RevisionColor}{HTML}{0000CD}
\newcommand{\ZF}[1]{}
\newcommand{\TS}[1]{}
\newcommand{\toref}[1]{}
\newcommand{\tocite}[1]{}
\newcommand{\todo}[1]{}

\newcommand{\nothing}[1]{}

\newcommand{\revv}[1]{{#1}}

\usepackage{acronym}
\newcommand{\code}{\texttt}
\newacro{AVX}[\code{AVX}]{Advanced Vector Extensions}
\newacro{AVX2}[\code{AVX2}]{Advanced Vector Extensions 2}
\newacro{BVH}{bounding volume hierarchy}
\newacro{CCD}{continuous collision detection}
\newacro{DOF}{degrees of freedom}
\newacro{EE}{edge-edge}
\newacro{FE}{finite element}
\newacro{FEA}{finite element analysis}
\newacro{FEM}{finite element method}
\newacro{FN}{false negative}
\newacro{FP}{false positive}
\newacro{GMP}{GNU Multiple Precision Arithmetic Library}
\newacro{GPU}{graphics processing unit}
\newacro{CPU}{central processing unit}
\newacro{HPC}{high-performance computing}
\newacro{IEEE}{Institute of Electrical and Electronics Engineers}
\newacro{IGA}{isogeometric analysis}
\newacro{IP}{incremental potential}
\newacro{IPC}{Incremental Potential Contact}
\newacro{MKL}{Intel Math Kernel Library}
\newacro{MSCCD}{minimum separation CCD}
\newacro{PE}{point-edge}
\newacro{PP}{point-point}
\newacro{PT}{point-triangle}
\newacro{TOI}{time of impact}
\newacro{VFX}{visual effects}
\newacro{TI}{Tight-Inclusion}
\newacro{SAP}{sweep and prune}
\newacro{BF}{brute force}
\newacro{FMA}{fused multiply-add}
\newacro{STQ}{Sweep and Tiniest Queue}
\newacro{ACCD}{Additive CCD}
\newacro{SH}{spatial hash}
\newacro{FRA}{fast root approximate}

\newcommand{\citet}[2][\textcolor{red}{XXX}]{#1~\cite{#2}}

\newcommand{\ssep}{\mid}
\newcommand{\M}{{
  \mathchoice{\mathcal{M}}{\mathcal{M}}{\scriptscriptstyle\mathcal{M}}{\mathcal{M}}
}}
\renewcommand{\S}{{
  \mathchoice{\mathcal{S}}{\mathcal{S}}{\scriptscriptstyle\mathcal{S}}{\mathcal{S}}
}}

\usepackage{calculator}
\newcommand{\numBroadmark}{12}
\ADD{\numBroadmark}{1}{\numOtherMethods} %
\ADD{\numOtherMethods}{1}{\numMethods} %
\ADD{\numMethods}{3}{\numImplementations} %

\title{Time of Impact Dataset for Continuous Collision Detection and a Scalable Conservative Algorithm
\funding{This work was funded by the NSF CAREER award under
Grant No. 1652515, the NSF grants OAC-1835712, OIA-1937043,
CHS-1908767, CHS-1901091, NSERC DGECR-2021-00461 and RGPIN 2021-03707, KAUST baseline funding (grant BAS/1/1679-01-01), by EU project DIGITbrain/ProMED
(952071). We thank the NYU IT High Performance Computing for resources, services, and staff expertise.}}%

\author{
David Belgrod\thanks{New York University (\email{db2762@nyu.edu}, \email{xz3752@nyu.edu}, \email{panozzo@nyu.edu}).} \and
Bolun Wang\thanks{KAUST (\email{bolun.wang@kaust.edu.sa}).}
\and
Zachary Ferguson\footnotemark[2] \thanks{Massachusetts Institute of Technology (\email{zfergus@mit.edu}).}
\and
Xin Zhao\footnotemark[2]
\and
Marco Attene\thanks{IMATI - CNR (\email{marco.attene@cnr.it}).}
\and
Daniele Panozzo\footnotemark[2]
\and
Teseo Schneider\thanks{University of Victoria (\email{teseo@uvic.ca}).}
}

\headers{TOI Dataset for CCD and a Scalable Conservative Algorithm}{Belgrod, Wang, Ferguson, Zhao, Attene, Panozzo, and Schneider}

\begin{document}

\maketitle

\begin{abstract}
We introduce a large-scale benchmark for broad- and narrow-phase \ac{CCD} over linearized trajectories with exact \aclp{TOI} and use it to evaluate the accuracy, correctness, and efficiency of {\numOtherMethods} state-of-the-art \acs{CCD} algorithms. Our analysis shows that several methods exhibit problems either in efficiency or accuracy.

To overcome these limitations, we introduce an algorithm for \acs{CCD} designed to be scalable on modern parallel architectures and provably correct when implemented using floating point arithmetic.
We integrate our algorithm within the \acl{IPC} solver \cite{Li2021CIPC} and evaluate its impact on various simulation scenarios.
Our approach includes a broad-phase \acs{CCD} to quickly filter out primitives having disjoint bounding boxes and a narrow-phase \acs{CCD} that establishes whether the remaining primitive pairs indeed collide.
Our broad-phase algorithm is efficient and scalable thanks to the experimental observation that sweeping along a coordinate axis performs surprisingly well on modern parallel architectures. For narrow-phase \acs{CCD}, we re-design the recently proposed interval-based algorithm of \citet[Wang et al.]{Bolun2021Benchmark} to work on massively parallel hardware.

To foster the adoption and development of future linear \acs{CCD} algorithms, and to evaluate their correctness, scalability, and overall performance, we release the dataset with analytic ground truth, the implementation of all the algorithms tested, and our testing framework.

\end{abstract}



\section{Introduction}

\begin{figure}
    \centering
    \includegraphics[width=\linewidth]{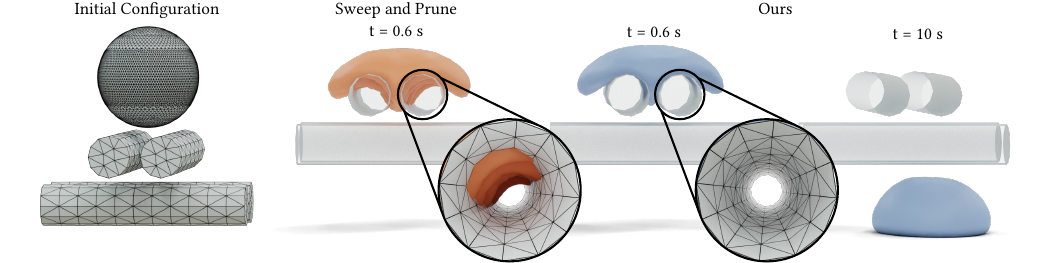}
    \caption{\emph{Approximate} collision detection can lead to poor simulation results. We simulate this scene using \acl*{IPC}~\cite{Li2020IPC} which provides an intersection-free guarantee, but approximate collision detection (either broad- or narrow-phase) can break this guarantee. Here we see using an approximate broad-phase method (\acl*{SAP}) results in missed collisions and intersections (see inset). In contrast, a conservative or exact method allows the ball to squeeze through the rollers and come out the bottom.}
    \label{fig:teaser}
\end{figure}

\Acf{CCD} is used extensively in engineering and scientific computing for the simulation of rigid and deformable objects. Objects are typically represented by triangle meshes. In this work, we focus on the common case where mesh vertices move along linear trajectories. Hence, collisions can occur either when an edge hits another edge or when a vertex hits a triangle~\cite{provot1997collision}.

\Ac{CCD} is usually divided into two steps: (1) broad-phase, which is a conservative filter that identifies candidate colliding pairs, and (2) narrow-phase, which validates each pair with an accurate and more computationally intensive algorithm. While the narrow-phase is local and involves only a pair of primitives, the broad-phase usually relies on acceleration data structures to prune unnecessary pairs and avoid the quadratic complexity of a brute-force evaluation on all possible pairs. Both problems have been extensively studied in graphics, engineering, and scientific computing in the last three decades (\cref{sec:related}).

An ideal linear \ac{CCD} algorithm takes the start and end points of a linear trajectory and determines if, at any point along the trajectory, the geometries intersect and, if they do, it tells us at which time the first collision happens. This might seem a trivial requirement for a \ac{CCD} algorithm; however, \citet[Wang et al.]{Bolun2021Benchmark} show that many popular \emph{narrow-phase} \ac{CCD} implementations are \emph{not} correct (\cref{fig:teaser}). Ensuring that an implementation is correct is a subtle and surprisingly difficult challenge since most of the intermediate computations use floating-point numbers, while the corresponding algorithms and their proof of correctness usually disregard floating-point rounding errors.

\paragraph{Two-Pronged Evaluation Approach}
Validating the correctness of algorithms implemented in floating-point is a major challenge, especially when multiple operating systems and architectures are considered. Indeed, floating-point computations might slightly differ due to either hardware specifics or modern compilers trying to reorder operations to improve performance. This makes it challenging to have correct code and especially hard to keep it up to date as compilers and architectures evolve. To provide a way to validate an implementation on a specific system, we have collected a large dataset of \ac{CCD} queries evaluated using exact computations (\cref{sec:dataset}), and we provide a large statistical experimental validation of several methods.
Differently from other CCD datasets \cite{Bolun2021Benchmark,Serpa:2020:Broadmark}, we generate \emph{ground truth \ac{TOI}} for each successive pair of frames from each scene, using a combination of symbolic computation and conservative filtering. This allows us to evaluate CCD as a whole, with broad and narrow phases coupled, and measure how conservative the methods are.

\paragraph{Evaluation}
We use exact computation to evaluate the correctness of {\numOtherMethods} broad phase algorithms on a large dataset (\cref{sec:comparison}). To our surprise, most of the implementations are incorrect (even if the methods described in the papers are when implemented with exact arithmetic) and miss collisions, making them unusable for interior point optimization.

\paragraph{Broad-Phase}
The broad-phase of a \ac{CCD} algorithm usually aims at detecting collision between (axis-aligned) boxes around primitives. Several existing methods simplify the problem by either checking for collisions only at the end of the time interval (i.e., discrete collision detection) or by assuming that only a small fraction of the scene moves. These simplifications allow for faster algorithms but may lead to unrealistic results when dealing with elastic bodies. Many acceleration data structures exist (e.g., hash grids, spatial trees, bounding volumes hierarchies) to avoid checking ``far away'' boxes, each providing different advantages in different situations. However, most of these structures are complex to parallelize, in particular on \acp{GPU} where dynamic memory allocation is not an option. Additionally, these structures' complexity makes it hard to verify and ensure correctness when using floating-point computations. In our work, we discovered that the simplest strategy, sweep along the most varying principal axis, is the most effective on \acp{GPU}. The algorithm requires only a parallel sort for the sweep, then every \ac{GPU} core will compare pairs of boxes. This strategy is not only massively parallel, but it is also trivial to ensure correctness: the only computation performed on the boxes is a comparison between floating-point numbers, which is exact.

\paragraph{Narrow-Phase}
\citet[Snyder et al.]{Snyder1993Interval} introduced a conservative narrow-phase \ac{CCD} algorithm that properly handles numerical errors by employing interval arithmetic. \Ac{TI} \ac{CCD}~\cite{Bolun2021Benchmark} used a similar idea and developed a faster \ac{CCD} method by replacing intervals with inclusion predicates. Unfortunately, \ac{TI} cannot be directly translated to \ac{GPU} as it contains several branches, dynamic allocation, and high register use\footnote{On modern GPU architectures, the number of registers per core is extremely limited, and this puts a limit on how many local variables can be used in every function.}. We propose a novel algorithm based on the same idea as \ac{TI}, redesigned to be \ac{GPU} friendly.

\paragraph{Contributions}
In our work, we introduce a dataset for \ac{CCD} on linearized trajectories of five scenes obtained from different simulators containing between 50 thousand to half a million primitives (i.e., vertex-face and edge-edge).  For every successive pair of frames, we compute the ground truth Boolean result and \ac{TOI} using a symbolic solver~\cite{Mathematica}. We use this dataset to validate the output of several \ac{CCD} algorithms.

Additionally, we introduce a novel \ac{GPU} \ac{CCD} implementation. Given two meshes for the start and end of a step, our method returns the time at which the impact occurs. Our pipeline includes the novel parallel broad-phase algorithm we call \emph{\ac{STQ}} as well as a GPU-friendly variant of the \ac{TI} algorithm. 

\revv{To foster the adoption and development of future linear \ac{CCD} algorithms, and to evaluate their correctness, scalability, and overall performance, we release the dataset with analytic ground truth, the implementation of all the algorithms, our testing framework, and the implementation of our \ac{STQ} and \ac{GPU} \ac{TI}. Links to all of these (and our accompanying video) can be found on our web page: \href{https://continuous-collision-detection.github.io/scalable_ccd}{continuous-collision-detection.github.io/scalable\_ccd}.}

\section{Related Work}
\label{sec:related}

We present an overview of existing CCD datasets and the broad- and narrow-phase collision detection algorithms benchmarked in our study. We refer to~\cite{Serpa:2020:Broadmark} for a detailed review of broad-phase algorithms and to \cite{Bolun2021Benchmark} for narrow-phase algorithms.

\subsection{Datasets}

The UNC Dynamic Scene Benchmarks~\cite{UNCBenchmark} features keyframes from simulation data and is commonly used throughout collision detection works as a source of benchmark data. This dataset covers a variety of simulation methods, materials (e.g., deformable and rigid), and physics (e.g., cloth and fracturing solids). We borrow three scenes (cloth-funnel, cloth-ball, and n-body simulation) from this dataset. We enrich these scenes with ground truth Boolean results and symbolic time of impacts which was not included in the original dataset.

\citet[Serpa and Rodrigues]{Serpa:2020:Broadmark} benchmark several classic broad-phase collision detection algorithms. In doing so, they provide not only reference implementations of these algorithms but also a benchmarking framework and procedurally generated scenarios. These benchmark scenes focus on simple primitives (e.g., cubes and spheres) in free fall or undergoing random rigid motion. In contrast, we focus on the more general case of  deformable triangle meshes. \citet[Serpa and Rodrigues]{Serpa:2020:Broadmark} focuses primarily on static collision detection, while our work evaluates broad-phase methods on continuous collision detection scenarios.

\citet[Wang et al.]{Bolun2021Benchmark} introduced a large scale dataset for narrow-phase \ac{CCD} algorithms. The dataset is designed to cover common cases extracted from simulation scenarios and challenging degenerate cases. \citet[Wang et al.]{Bolun2021Benchmark} uses the dataset to evaluate the accuracy (the number of false positives), correctness (the number of false negatives), and efficiency (the average runtime) of different narrow-phase \ac{CCD} algorithms. However, the large-scale dataset proposed by \citet[Wang et al.]{Bolun2021Benchmark} contains only the queries and the ground truth Boolean results, but not the collision time for each query. We propose a new benchmark dataset of over 4M collisions combined with their \ac{TOI} (\cref{sec:dataset-toi}). This allows us to evaluate the accuracy of different methods in addition to their correctness.

\subsection{Broad-Phase}
We discuss the {\numBroadmark} methods benchmarked in~\cite{Serpa:2020:Broadmark} and additionally include the spatial hash data structure used in~\cite{Li2020IPC}. We note that broad phase methods can be used for both static or continuous collision detection by just changing the geometric proxies used: instead of building a proxy, such as a bounding box, around a static object, it is possible to build the proxy around the linearized trajectory in a time-step. Their performance is, however, very different in these two settings due to the much larger number of overlaps in the continuous collision detection case.

\paragraph*{BF}
A simple \emph{CPU parallel} \emph{brute-force} check where every box in the list is checked against every other box. To avoid any concurrent accesses to the resulting candidates, we use a synchronized vector~\cite{Dagum:1998:OpenMP}, and the algorithm is accelerated using \ac{AVX} instructions. This algorithm is simple, but its complexity is $O(n^2)$, with $n$ the number of primitives: it is not practical for large scenes, but it is viable for smaller ones.

\paragraph*{SAP} A serial and parallel standard implementation of the \emph{sweep and prune algorithm} \cite{Baraff:1992:Dynamic,Capannini2016Adaptive,Capannini2016Succinct,CapanniniAdaptive2018}: it starts by performing an intersection check between the $x$-axes of the boxes by sorting the boxes along $x$. For every $x$-intersecting box, the algorithm proceeds by checking $y$ and $z$, every time sorting and pruning the axis. This algorithm improves over the simple brute force as its complexity is $O(n\log(n))$.
SAP comes also with an OpenCL \ac{GPU} implementation from Bullet 3~\cite{Coumans:2019:Bullet} based on~\cite{liu:2010:real}. Note that we cannot provide a full comparison with the reference implementation because of the incomplete state of their code.

\paragraph*{iSAP} A serial implementation of \emph{incremental sweep and prune algorithm} \cite{Coumans:2019:Bullet}. It is an improvement over SAP: For every intersecting box along the $x$-axis, the pair of overlapping boxes is added to a list.
Three lists are built to keep track of intersecting boxes along all three axes: $x$, $y$, and $z$. Finally, a pass through all three arrays is done to find pairs of boxes that intersect along all three axes. 

\paragraph*{Grid} The scene is divided into voxels (or cells) of uniform size $v$. The voxel size $v$ is chosen based on a target number of boxes per voxel (we use the default: 200 boxes per voxel). Every input bounding box is assigned to the cells intersecting it. We detect intersections between boxes by iterating over the sparse cells. This algorithm is efficient and easily parallelizable (on both CPU and GPU); however, its main disadvantage is the choice of $v$: a small $v$ will lead to many boxes, and an explosion in memory due to duplicate collision candidates, and a large $v$ leads to many boxes inside each cell (e.g., if there is only one voxel the grid reduces to \ac{BF}). The choice of $v$ is particularly problematic for large displacements. For instance, if the objects are moving apart, the grid (and the number of boxes) will grow in size, potentially leading to an exploding number.
For the GPU algorithm, we use the OpenCL parallel implementation based on Bullet~\cite{Coumans:2019:Bullet}. We note that it requires the same delicate choice of $v$ that is even exacerbated as GPUs have less memory than CPUs.

\paragraph*{GSAP} This method is similar to the Grid method, but uses a \ac{SAP} inside each cell instead of a brute force check. It suffers from similar issues as Grid: A small $v$ will result in many duplicate candidates and large memory usage, but, unlike Grid, a large $v$ reduces to \ac{SAP} and is, therefore, more efficient than Grid's \ac{BF} comparisons. 

\paragraph*{SH} A parallel CPU implementation of Grid, that encodes the grid implicitly using a \emph{spatial hashing} function~\cite{Tang:2018:I-Cloth,Li2020IPC,Tang:2018:PSCC}. Each candidate box is rasterized in the grid, and for each voxel, a hash value is computed. These hash values are used to store the elements IDs in a hash map (mapping from voxel indices to a vector of element IDs contained inside the voxel). The candidate collisions can then be found by rasterizing the query element and looking up the voxel indices. Our implementation is based on the code from~\cite{Li2020IPC}, which has been modified to produce all collision candidates in one parallel loop and include axis-aligned bounding box checks of elements, to make it comparable with the other approaches. This algorithm has the same shortcomings as Grid: a wrong choice of $v$ might lead to either slow performances or excessive memory usage. We use a heuristic for the voxel size equal to two times the maximum of the average edge lengths and the average displacement length. This ensures the average element fits within a single voxel.

\paragraph*{BVH} A \emph{bounding volume hierarchy}. The boxes are divided into two sets: query and target. The target queries are sorted using Morton encoding to optimize spatial locality. The sorted boxes are grouped into pairs, each forming a larger box. By recursively iterating the process, we obtain a \emph{binary} tree, where the root is a box containing the whole space-time scene. Every query box traverses the tree by recursively checking its intersection with the box at the tree's node until it reaches the leaf. The BVH can be updated ``bottom-up'' (i.e., if a leaf box grows, it can update its parent until the root), dramatically reducing the update cost in dynamic scenes.
We use a \emph{deferred} BVH \emph{(DBVT-D)}, which performs a single tree-tree query.
Additionally, \citet[Bridson et al.]{Bridson:2002:Robust} proposes to use numerical tolerances to account for rounding error in floating-point computation, which we also apply to all methods we compare against to mitigate (but as we will show in \cref{sec:comparison} not reliably address) the effect of rounding errors.
We also include an OpenCL GPU implementation of the Linear BVH in Bullet~\cite[Coumans and Bai]{Coumans:2019:Bullet}. Similar to the BVH, the tree is organized using Morton encoding. By default, \citet[Serpa and Rodrigues]{serpa:2019:flexible} assumes a maximum of $18n$ possible intersections, with $n$ input bounding boxes, and discards any successive one. Changing the default size for all our scenes required a trade-off between performance for smaller scenes in exchange for more collisions in larger scenes. To avoid this unnecessary shortcoming, which introduces false negatives, we changed the algorithm to process all intersections in an appropriate amount of time: If the list of candidates reaches the maximum, we stop storing them and count their number. Once we know the total number of candidates, we re-execute the algorithm with the correct preallocated size. This change affects only a few cases where the number of candidates exceeds $18n$.

\paragraph*{KDT} An optimized \emph{KD-Tree} designed to handle static scenes based on the efficient implementation in~\cite{serpa:2019:flexible}. The spatial subdivision is designed to adaptively partition the space and have a small number of boxes attached to each cell. We note that when using the automatic box inflation present in the implementation~\cite{serpa:2019:flexible}, the algorithm does not report any collision (i.e., it fails to detect true positives). We thus alter the code to disable this feature and use our default 1\% inflation, which leads to reasonable results  suggesting a bug in the auto inflation code.

\paragraph*{Tracy} The parallel method of~\citet[Tracy et al.]{tracy:2009:efficient}, which builds off the incremental SAP of~\citet[Baraff]{Baraff:1992:Dynamic} and~\citet[Cohen et al.]{Cohen:1995:ICOLLIDE} by including the ability to insert AABBs without the need to perform a full sort of the axes.

\paragraph*{CGAL} The CGAL implementation of an interval-tree SAP algorithm designed to handle $d$-dimensional axis-aligned boxes (in our experiments, we use $d=3$)~\cite{Zomorodian:2000:Fast,cgal}. This method works by using SAP on the first axis and then using range and interval trees on the subsequent axes.

\subsection{Narrow-Phase}

\paragraph{Numerical Root-Finding} Narrow-phase CCD can be reduced to root-finding: the roots of a carefully designed function correspond to the \ac{TOI}(s). CCD of linear trajectories without minimal separation equates to finding the roots of a cubic polynomial~\cite{provot1997collision}. Many methods focus on solving these cubic polynomials using numerical methods~\cite{provot1997collision,Yuksel2022High}. \citet[Provot]{provot1997collision} introduces the most common strategy of finding a time of coplanarity and then performing an inside check. This idea has since been expanded to solve both rigid~\cite{Redon2002fast,kim2003collision} and deformable collisions~\cite{Bridson:2002:Robust,Hutter2007optimized,Tang2011VolCCD}.

\revv{\citet[Lan et al.]{Lan2022Penetration} recently proposed the \ac{FRA} \ac{CCD} method which breaks down the cubic equation into three monotonic intervals, identifies which interval the \ac{TOI} may reside in, and uses an altered Newton-Raphason search to find an approximate root smaller than the actual \ac{TOI}. We briefly discuss this method further in \cref{sec:np-comparison}.}

The downside of these methods is that they assume infinite precision. When implemented using floating-point numbers, these methods can both miss collisions (false negatives) and report non-existent collisions (false positives).

\paragraph{Inclusion-Based Root-Finding} As an alternative to numerical root-finding algorithms, some propose using inclusion-based root-finding algorithms to determine if a root exists in the co-domain of our function with some tolerance~\cite{snyder1992interval,Snyder1993Interval,Redon2002fast,brian1990geometric,Bolun2021Benchmark}. This can either be done using interval arithmetic~\cite{snyder1992interval} or by designing custom inclusion functions~\cite{Bolun2021Benchmark}. The latter has the benefit of producing tighter inclusion functions than general interval arithmetic and can be performed in floating-point with specially crafted error bounds. These methods avoid false negatives but produce false positives, which add extra numerical padding to simulated objects and can result in worse convergence when used in line-search-based implicit solvers.

\paragraph{Conservative Advancement}
Originally introduced for CCD between rigid convex objects~\cite{Mirtich1996}, conservative advancement incrementally estimates the \ac{TOI} through a series of minimum distance queries. This work has been subsequently extended to non-convex~\cite{Zhang:2006:Interactive}, articulated~\cite{Zhang2007Continous}, and polygon-soup models~\cite{Tang2009C2A}. Most notably in the context of this work, \citet[Tang et al.]{Tang:2010:Continuous} proposed a method of ``Local Advancement'' for CCD between deformable triangles.
\citet[Li et al.]{Li2021CIPC} propose a conservative advancement method for CCD between triangles. They claim numerical robustness as they attempt to conservatively underestimate the \ac{TOI}. As this work postdates the benchmark and analysis done by~\citet[Wang et al.]{Bolun2021Benchmark}, we investigate here their claims and determine if it is a viable alternative to the \ac{TI} algorithm of~\citet[Wang et al.]{Bolun2021Benchmark}. As shown in \cref{sec:np-comparison}, the 
\revv{method can miss collisions as one requests a more accurate \ac{TOI}.
This inaccuracy is due to floating-point rounding errors in distance computations (involving square roots), which make the CCD algorithm in \citet[Li et al.]{Li2021CIPC} not conservative and potentially unsuitable for interior point optimization.}

\paragraph{Exact Methods} Both \citet[Brochu et al.]{brochu2012efficient} and \citet[Tang et al.]{tang2014fast} introduce exact root-parity CCD methods. Root parity is insufficient for CCD, as it cannot distinguish between 0 or 2 collisions within a timestep. In addition, their algorithm is not handling certain corner cases, making the root parity check not exact \cite{Bolun2021Benchmark,Wang2022RootParity}. 

The only method exact CCD method we are aware of is using symbolic root finding~\cite{Mathematica}. Unfortunately, this is computationally expensive (seconds for each query) and thus impractical in a simulation setting. We use the ground truth generated by this method to verify the results of all queries used in this work.

\section{Preliminaries}

We formally introduce the CCD problem and classifications of \ac{CCD} algorithms, which we will use in our evaluation. A CCD algorithm considers a scene with objects (typically triangles) moving from an initial position at time $t=0$ to a final position at time $t=1$. If two objects intersect during the movement, the algorithm should report \texttt{COLLISION} and return the time $t^\star \in [0,1]$ at which such a collision occurs ($t^\star = \infty$ if there are no collisions). As mentioned, algorithm implementations may not be robust. If an algorithm reports \texttt{COLLISION} but no pair of objects actually intersect, we call the report a \emph{false positive}. When an intersection actually occurs, but the algorithm does not report \texttt{COLLISION}, we call the report a \emph{false negative}.

\begin{definition}
A CCD algorithm (or implementation of a CCD algorithm) is \emph{exact} if it reports \texttt{COLLISION} if \emph{and only if} the geometries intersect.
\end{definition}

Hence, an exact algorithm never reports false positives and/or negatives.
This property can be achieved using exact or symbolic calculations as explained in \cref{sec:dataset}, though this is intractably slow in real applications. Nonetheless, most algorithms can tolerate few false positives; thus, we introduce a new definition.

\begin{definition}
A CCD algorithm (or an implementation of a CCD algorithm) is \emph{conservative} if, when the geometries intersect, it reports \texttt{COLLISION}.
\end{definition}

Therefore, a conservative algorithm never reports false negatives but may report false positives.
This property is fundamental when false negatives cannot be tolerated, such as in a new trend of contact models and corresponding simulators \cite{Li2020IPC,Ferguson2021RigidIPC,Li2021CIPC} which guarantee (and also assume), by construction, that no interpenetrations are present in the scene at any moment in time. These algorithms provide robust and accurate modeling of contact but are unable to tolerate \ac{CCD} imprecision: if a \ac{CCD} query misses a collision, a self-intersection will appear, breaking the assumption of a self-intersection-free state and thus prevents the simulation from terminating (\cref{fig:teaser}). This imprecision includes floating-point rounding errors, which need to be accounted for in a conservative \ac{CCD} algorithm. For this class of algorithms, we measure their \emph{accuracy} (i.e., we measure how close they are to be exact) by counting the number of false positives they produce.

Not all contact models require conservative \ac{CCD}\revv{~\cite{Chen2023Shortest,Macklin2016XPBD,verschoor2019efficient}}. In some cases, for instance, the model introduces contact forces to remove intersections between primitives. Since the forces are introduced after a collision to remove the intersections,
small numerical errors in the \ac{CCD} queries are unlikely to affect the overall simulation, and it is thus common to sacrifice numerical guarantees in the \ac{CCD} algorithm, favoring approximations that lead to a lower computation cost.
\begin{definition}
A CCD algorithm (or an implementation of a CCD algorithm) is \emph{approximate} if, most of the time, it reports \texttt{COLLISION} when the geometries intersect; however, some collisions may be missed.
\end{definition}
In this case, it is important to measure both false positives (to see how close to exact they are) and false negatives (to ensure that only a few collisions are missed).

\section{Dataset Generation}\label{sec:dataset}

\begin{figure}\scriptsize\centering
    \parbox{.17\linewidth}{Armadillo-Rollers}\hfill
    \parbox{.16\linewidth}{\centering Cloth-Ball}\hfill
    \parbox{.16\linewidth}{\centering Cloth-Funnel}\hfill
    \parbox{.16\linewidth}{\centering N-Bodies}\hfill
    \parbox{.16\linewidth}{\centering Rod-Twist}\hfill
    \parbox{.16\linewidth}{\centering Puffer Ball}\\
    \parbox{.16\linewidth}{\centering\includegraphics[width=\linewidth]{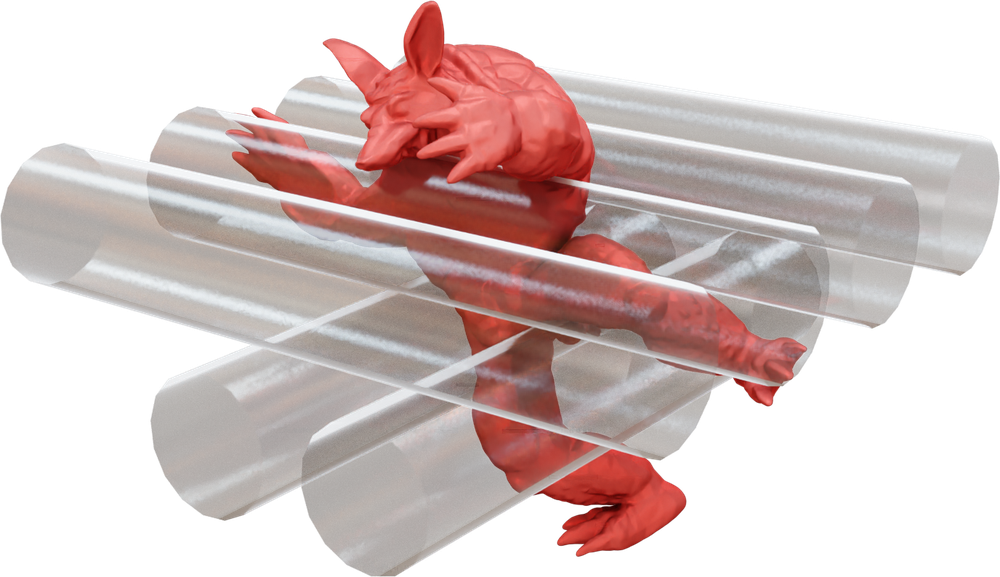}}\hfill
    \parbox{.16\linewidth}{\centering\includegraphics[width=\linewidth]{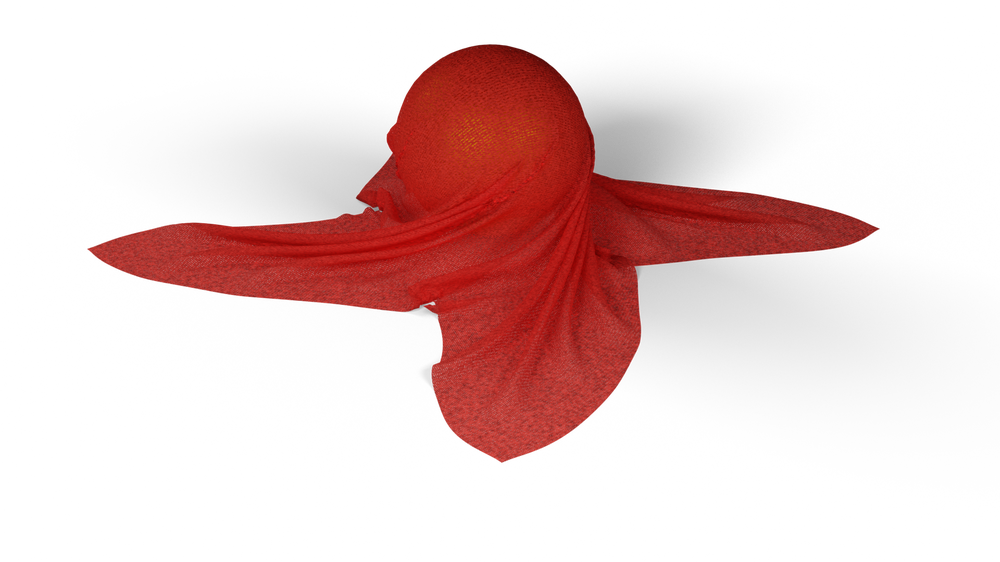}}\hfill
    \parbox{.16\linewidth}{\centering\includegraphics[width=\linewidth]{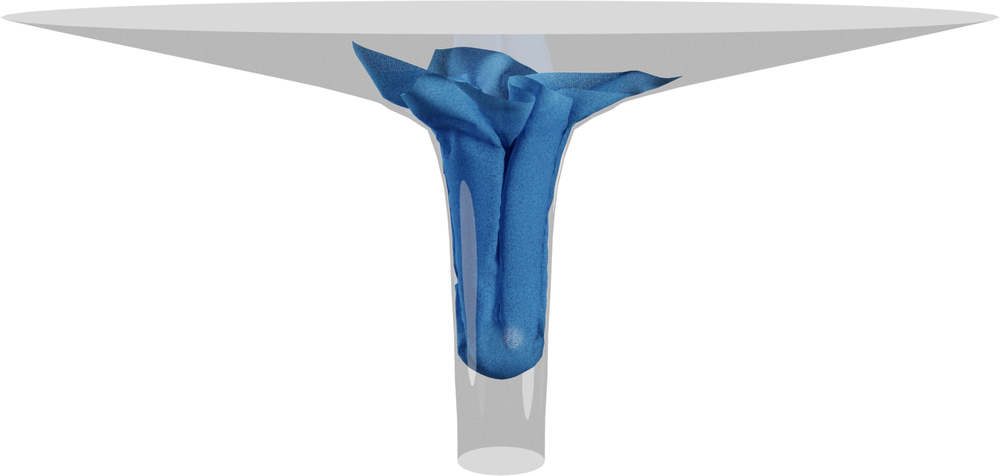}}\hfill
    \parbox{.16\linewidth}{\centering\includegraphics[width=\linewidth]{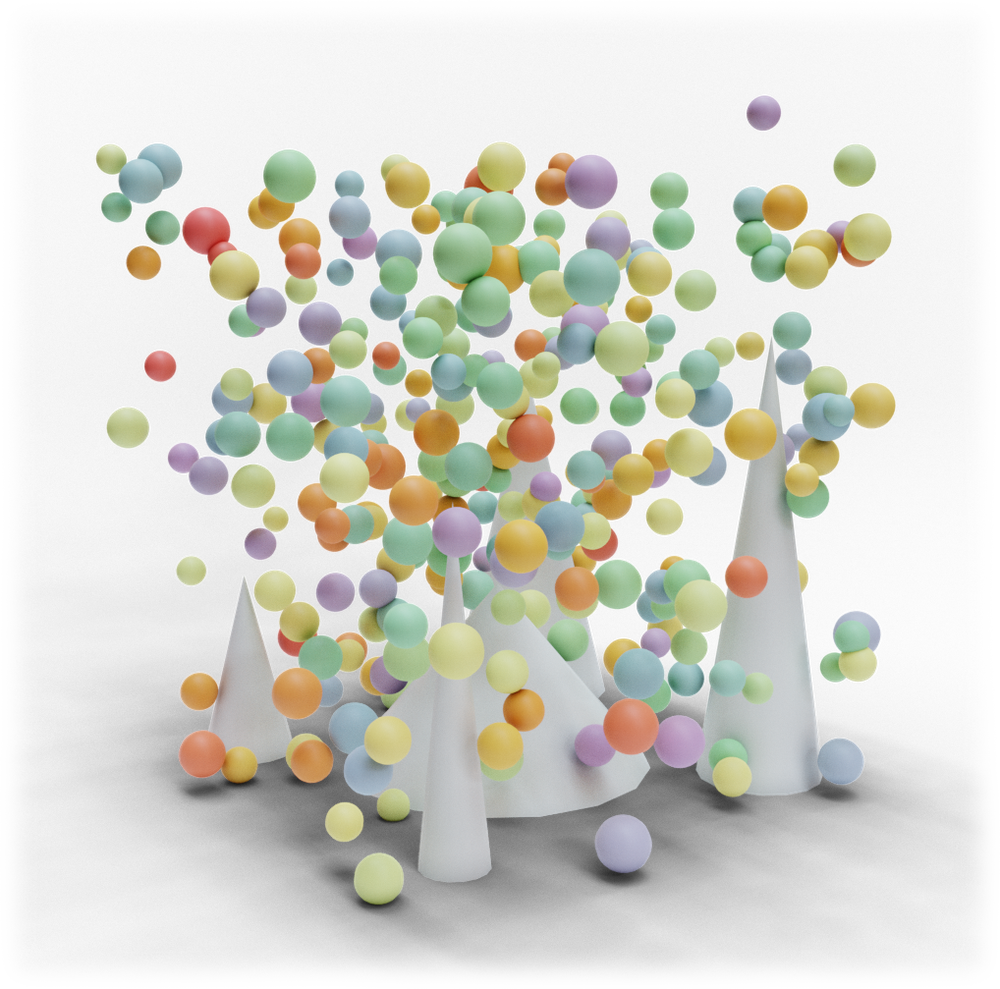}}\hfill
    \parbox{.16\linewidth}{\centering\includegraphics[width=\linewidth]{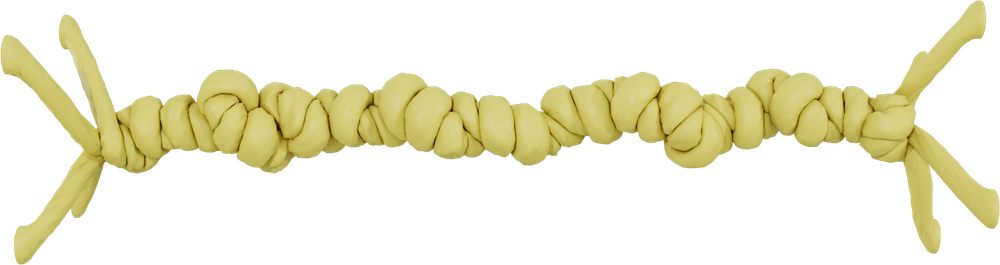}}\hfill
    \parbox{.16\linewidth}{\centering
    \includegraphics[width=\linewidth]{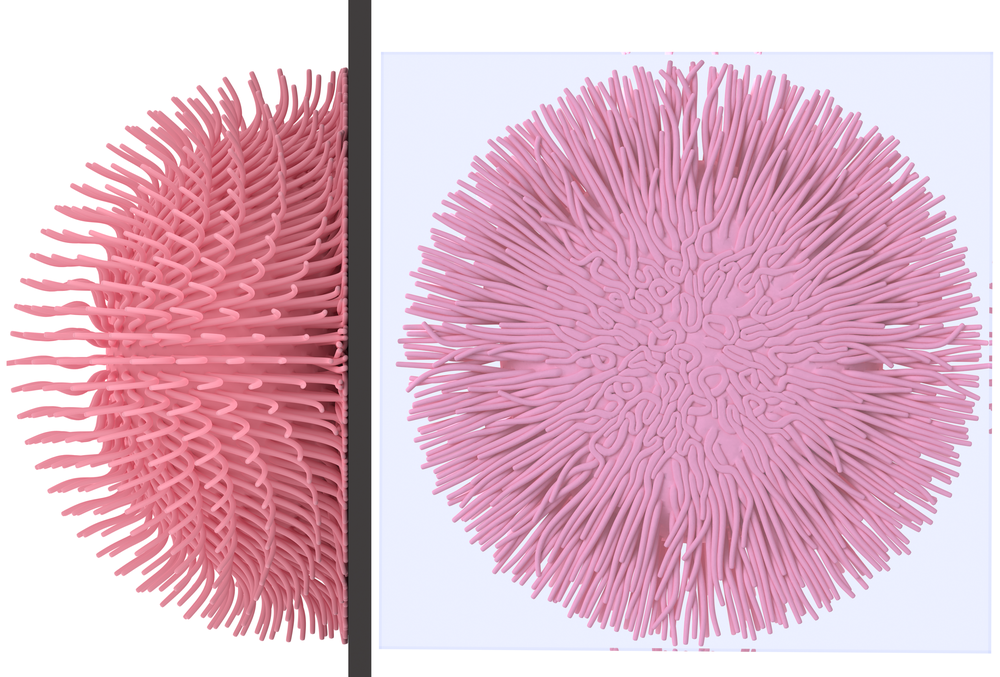}
    }\par
    \parbox{.16\linewidth}{$|B|=72{,}740$}\hfill
    \parbox{.16\linewidth}{\centering $277{,}653$}\hfill
    \parbox{.16\linewidth}{\centering $55{,}864$}\hfill
    \parbox{.16\linewidth}{\centering $440{,}280$}\hfill
    \parbox{.16\linewidth}{\centering $239{,}600$}\hfill
    \parbox{.16\linewidth}{\centering $3{,}192{,}650$}\par
    \caption{Overview of the six scenes in our dataset. The bottom row indicates $|B| = |V| + |E| + |F|$ (i.e., the number of boxes) where $|V|$, $|E|$, and $|F|$ are the number of vertices, edges, and faces, respectively.}
    \label{fig:dataset}
\end{figure}

Our dataset is composed of \revv{six} simulated scenes (\cref{fig:dataset}) and the corresponding ground truth data for continuous collisions between frames. From the UNC Dynamic Scene Benchmark~\cite{UNCBenchmark}, we include two codimensional cloth simulations with a large number of self-collisions (cloth-ball and cloth-funnel) and a simulation of a large number of rigid spherical bodies (n-bodies). We also include \revv{three} elastodynamic scenes featuring large compression and nonlinear buckling simulated using the \ac{IPC} method~\cite{Li2020IPC} (armadillo-rollers, rod-twist, \revv{and puffer ball}).

\revv{These scenes were chosen because they highlight a variety of dynamics (both rigid and soft body) and element types (codimensional shells and volumetric elements). In comparison to the UNC Dynamic Scene Benchmark~\cite{UNCBenchmark}, we include scenes with larger deformations (rod-twist) and larger element counts (puffer ball: 3.2M boxes). In comparison to \citet[Wang et al.]{Bolun2021Benchmark}, we include higher-resolution scenes\footnote{\revv{E.g., our rod-twist is 240K boxes versus \citet[Wang et al.'s]{Bolun2021Benchmark} mat-twist is only 19K boxes, or our puffer ball is 3.2M boxes versus \citet[Wang et al.'s]{Bolun2021Benchmark} golf ball is 124K boxes.}} with more complex deformations and challenging dynamics. In comparison to \citet[Serpa and Rodrigues]{Serpa:2020:Broadmark}, we focus our benchmark more on elastodynamics with continuous collisions between triangle meshes rather than discrete collisions between individual boxes under free-fall, Brownian motion, or gravity.}

\paragraph*{Ground Truth} We generate ground truth for each successive pair of frames from each scene, using a combination of symbolic computation and conservative filtering. While this might seem similar to the dataset of narrow-phase ground truth introduced by \citet[Wang et al.]{Bolun2021Benchmark}, they only provide a set of queries with no global mesh, nor are the queries separated by time step. Therefore, the dataset of \citet[Wang et al.]{Bolun2021Benchmark} is only suited for evaluating narrow-phase CCD algorithms, and this necessitates introducing a new dataset to evaluate broad-phase methods or CCD as a whole.

We first enumerate all possible collision pairs (collision candidates) through a brute-force approach. We only consider point-triangle and edge-edge pairs as these pairs capture the first collisions between triangles~\cite{provot1997collision} (not including points that are vertices of the triangles and edges that share a common endpoint). For each collision candidate, we determine if the pair collides using the provably conservative CCD of~\citet[Wang et al.]{Bolun2021Benchmark}. While this CCD algorithm is guaranteed to not have false negatives, it may produce false positives. To eliminate false positives, we find the exact Boolean solutions of the CCD query using the symbolic solver in Mathematica~\cite{Mathematica}. Mathematica uses symbolic computations combined with exact arithmetic to produce a symbolic expression for the \ac{TOI}. We use Mathematica’s exact predicate computation to determine if there is a valid \ac{TOI}  ($t\in[0, 1]$) and, therefore, a collision.

While we could skip the middle step and directly use the symbolic CCD, this would be prohibitively slow, as the symbolic solvers take several seconds per query. Instead, the method of~\citet[Wang et al.]{Bolun2021Benchmark} quickly filters the majority of the collision pairs, leaving a smaller number of candidates to validate with the symbolic solver.

\paragraph{Time of Impact Expressions}\label{sec:dataset-toi}
In addition to the Boolean ground truth, our dataset is the first to include the symbolic expressions for the valid roots (as Wolfram Language \revv{WXF} files). It is necessary to save the symbolic expressions (instead of a real or rational number) as they may include operators and functions that cannot be evaluated exactly using floating point or rational numbers (e.g., square roots). For example, \cref{eq:root-example} shows that Mathematica stores the \ac{TOI} as the roots of a cubic polynomial which can be solved analytically (e.g., by using Cardano's method~\cite{Cardano_1545}) but requires irrational and complex arithmetic.

\begin{equation}\label{eq:root-example}
\begin{split}
t^\star = \operatorname{Root}(32438097225180438401964874473761976929223\, t^3\\
-~2546681458666122439357688750343089434416006\, t^2\\
+~2471187477357729479707126378291971630585896\, t\\
-~266132451806881357156163391768279366354706, 2)
\end{split}
\end{equation}

In total, this includes over
3.3M edge-edge %
and 728K vertex-face contacts. %
For reproducibility, we provide scripts to rerun the validation on other architectures and compilers.

\section{Algorithm}\label{sec:method}

\begin{algorithm}
\caption{Overview of our CCD algorithm.}
\label{alg:overview}
\begin{algorithmic}[1]
\Function{CCD}{$M_0, M_1$}
    \State $B \gets $\Call{BuildBoxes}{$M_0, M_1$} \Comment{CPU, \cref{sec:m:boxes}}
    \State $C \gets $\Call{BroadPhase}{$B$} \Comment{GPU, \cref{sec:m:bp}}
    \State $t^\star \gets $\Call{NarrowPhase}{$C, M_0, M_1$} \Comment{GPU, \cref{sec:m:np}}\label{line:overview-np}
    \State \Return $t^\star$
\EndFunction
\end{algorithmic}
\end{algorithm}

Our CCD algorithm for triangle meshes, summarized in \cref{alg:overview}, takes as input two triangle meshes $M_0, M_1$ whose vertex positions are given at time $t=0$ and $t=1$ and are assumed to move along linear trajectories in between. The algorithm returns the earliest \ac{TOI} $t^\star$ ($t^\star = \infty$ if there are no collisions between $M_0$ and $M_1$). Vertex coordinates are represented using floating-point numbers.
Our parallel algorithm is implemented both on GPU and CPU and strives to vectorize the computations. 

Our broad and narrow phase algorithms are inspired by the results (\cref{sec:comparison}): for broad phase, the SAP algorithms are among the fastest and simplest (we exclude SH as its runtime depends on the grid size, \cref{fig:sh-voxels}), and only TI is conservative for the narrow phase. Thus we develop our algorithm on their core principle while redesigning it to target vectorized architectures and flexible memory management.

\paragraph{Overview}
We first build a set $B=\{b_i \ssep i=1,\dots,k\}$ of $k$ boxes $b_i=(b_i^m, b_i^M)$ (where $b_i^m$ and $b_i^M$ are the minimum and maximum corner of the box respectively) on CPU around every moving primitive (triangles, edges, and vertices at both $t=0$ and $t=1$) on $M_0$ and $M_1$. We use the CPU for this step as its running time is negligible. Additionally, we need to have access to the whole scene that might not fit on GPU.
When moving the boxes to GPU, we represent them in \emph{single} precision while ensuring that every input (in double precision) is \emph{exactly} contained in its box $b_i$ (\cref{sec:m:boxes}). This phase is fast and takes around 10\% of the runtime (\cref{fig:our-timings}).

Then we pass $B$ to our broad phase algorithm to discard any far away candidate (\cref{sec:m:bp}). The broad phase produces a set of $n$ candidates intersection pairs $C=\{(l_i, r_i) \ssep 1\le i \le n\}$, where every pair $(l_i, r_i)$ indicates that the primitives in the boxes $b_{l_i}$ and $b_{r_i}$ potentially intersect. This stage takes about 50\% of the runtime (\cref{fig:our-timings}). The main idea of our algorithm is similar to Sweep and Prune~\cite{Baraff:1992:Dynamic,Cohen:1995:ICOLLIDE}, but we vectorize the operation and avoid assumptions on the number of intersections to avoid memory allocation. In particular, SAP~\cite{liu:2010:real} uses the length of the box as a heuristic to determine the number of threads used to process a box; instead, we use a \emph{consumer queue} which keeps the work balanced.
While there are similarities between SAP and our algorithm, our version is conservative and is $3.76\times$ faster on average (\cref{fig:stats}).

Finally, to obtain the \ac{TOI} $t^\star$, we run our narrow phase algorithm using the collision candidates $C$, and the input meshes $M_0$ and $M_1$ (\cref{sec:m:np}). The core idea is the same as in~\cite{Bolun2021Benchmark}, but we redesigned the algorithm to avoid recursion and used a \emph{worker queue} paradigm to make it GPU parallelizable.
In comparison, we implemented~\cite{Bolun2021Benchmark} in CUDA and observed that it does not scale at all with multiple threads, due to branching, and high-registry use, making it slower than its parallel CPU counterpart implemented by~\citet[Wang et al.]{Bolun2021Benchmark}.

On both CPU and GPU, we execute the narrow-phase using the same precision as the input (e.g., double precision for all our experiments). \revv{Alternatively, we could utilize single-precision floating-point numbers by adjusting our forward error analysis~\cite{attene20} to produce a larger $C_\epsilon$ in \cref{alg:np}. However, this results in a larger number of false positives. Therefore, we stick with double-precision.\footnote{Modern consumer GPUs have very limited support for double computation, but this is not an issue for our purposes, as the narrow-phase is memory bound and the lower number of double-precision ALUs does not affect the algorithm performance.}} 

In \cref{sec:guarantees} we show that our overall algorithm is \emph{conservative}; that is, it never misses collisions. \Cref{sec:comparison} shows that our algorithm has similar performance as state-of-the art and \cref{sec:results} details the performance of our method.

\subsection{Construction of the Boxes}\label{sec:m:boxes}
To construct a \emph{tight} single precision box $b=(b^m, b^M)$ around a primitive (i.e., a triangle, edge, or vertex), we first compute the extent of the box $b'^m, b'^M$ in double precision using the min and max of the coordinates of the primitive during its entire movement. Because trajectories are linear, considering $t=0$ and $t=1$ is sufficient. For instance, for a triangle, $b'^m$ is the minimum of the $x,y,z$-coordinates of the three vertices, each at either $t=0$ or $t=1$.
To \emph{conservatively} convert $b'^m$ in single precision, for each coordinate (e.g., $x$), we first round $b'^m_x$ to its nearest single precision value and check if $b^m_x < b'^m_x$, in case it is not, we decrease $b^m_x$ to its previous representable single precision value using the function \texttt{nextafterf}\footnote{\url{https://en.cppreference.com/w/c/numeric/math/nextafter}}. The procedure for $b^M$ is similar.
When running our algorithm on GPU, we need to copy the boxes on the device; in our experiments, this time (B2G) is negligible (\cref{fig:our-timings}).

\subsection{Broad-Phase}\label{sec:m:bp}
Our Sweep and Tiniest Queue (STQ) algorithm is based on the Sweep and Prune algorithm~\cite{Baraff:1992:Dynamic,Cohen:1995:ICOLLIDE}. We observe that on modern architectures, due to their memory layout and a large number of ALUs that favor heavier computation with structured memory access,  brute-force checking all possible pairs is not only easy to parallelize but extremely fast. Unfortunately, this simple approach has a runtime quadratic with respect to the number of pairs and cannot be applied to large scenes. To overcome this limitation, we borrow ideas from the SAP to limit the average complexity of our algorithm (\cref{fig:complexity}).

\paragraph{Algorithm}

\begin{algorithm}
\begin{algorithmic}[1]
    \Function{BroadPhase}{$B$}
    \State $B_C\gets \{c = (b^m + b^M)/2 \ssep b \in B\}$ \Comment{Box centers}
    \State $\sigma \gets \sigma(B_C)$ \Comment{Variance of the boxes centers}
     \State $a \gets \argmax_{i\in\{x,y,z\}} \sigma^i$  \label{line:variance}
     \State $a^c \gets \{x,y,z\} \setminus a$
     \State $B' \gets $\Call{sort}{$B$, \textproc{order}$_a$}\Comment{in parallel}
     \State $Q\gets \{\}$
     \ForAll{$i\in \{1, \ldots, |B'|-1\}$}\label{line:queue-init}
       \If{$b_{B'_i}^a\cap b_{B'_{i+1}}^a \neq \emptyset$} \Comment{Boxes intersect along $a$}
            \State $Q\gets Q\cup (b_{B'_i}, b_{B'_{i+1}})$
       \EndIf
     \EndFor
     \State
     \While{$Q\neq \emptyset$}
     \State $Q'\gets \{\}$
        \ForAll{$(b_i, b_j) \in Q$}
            \If{$b_{i}^{a^c}\cap b_{j}^{a^c} \neq \emptyset$} \label{line:inter-other} \Comment{Boxes intersect in other dimensions}
                \State $C\gets C \cup (b_i, b_j)$
            \EndIf
            \If{$b_{i}^a\cap b_{B'_{j+1}}^a \neq \emptyset$} \Comment{Next sorted box intersects along $a$}
                \State $Q'\gets Q' \cup (b_i, b_{B'_{j+1}})$\label{line:append-q}
            \EndIf

        \EndFor
     \State $Q\gets Q'$
     \EndWhile
    \State \Return $C$
\EndFunction
\State
\Function{order$_a$}{$b_i, b_j$}
    \State \Return $b_i^{m_a} < b_j^{m_a}$ \Comment{Order by min value along $a$}
\EndFunction
\end{algorithmic}
    \caption{Overview of the broad-phase.}
    \label{alg:bp}
\end{algorithm}

Our STQ (\cref{alg:bp}) starts by computing the variance $\sigma$ of the boxes' centers $B_C$ and finding the most varying axis $a$ (line~\ref{line:variance})~\cite{liu:2010:real}. We then sort the boxes $B$ along the axis $a$ based on their minimum $a$ coordinate and initialize a queue $Q$ that will hold pairs of boxes that overlap on the $a$-axis. Since $a$ is the axis of maximum variance, this will lead to the smallest possible queue among the three axes if the data is uniformly distributed.

In the first step, for every box $b_i$ we check if it intersects its next box $b_{i+1}$ along the $a$-axis; if it does, we append the pair $(b_i, b_{i+1})$ to $Q$ (line~\ref{line:queue-init}).

\begin{figure}
    \centering
    \includegraphics[width=3.5in]{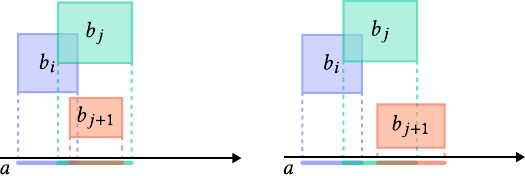}
    \caption{The boxes $b_i$, $b_j$, and $b_{j+1}$ are sorted along the axis $a$, \revv{and their indices indicate the order along $a$.} Since $b_i$ and $b_j$ intersect, we append the pair to $C$ and check $b_{j+1}$. If $b_{j+1}$ intersects $b_{i}$ on the axis $a$ (left) we append the ($b_{i}$, $b_{j+1}$) to $Q'$, if not (right) we discard the pair. Because the boxes are sorted along the axis $a$, if $b_{j+1}$ does not overlap with $b_i$, then $b_k$ for all $k > j+1$ do not intersect $b_i$, and we can skip checking them.}
    \label{fig:bp}
\end{figure}

Then STQ extracts a pair $(b_i, b_j)$ from $Q$ and checks if it intersects in the remaining two axes $a^c$ (line~\ref{line:inter-other}). If they do, we append the pair to the output $C$ (global). Finally we add the pair $(b_i, b_{j+1})$ to an output queue $Q'$ if $(b_i, b_{j+1})$ intersects along the $a$-axis (line~\ref{line:append-q} and \cref{fig:bp}).

\paragraph{Implementation Remarks} We remark that the STQ algorithm creates queues with \emph{at most} $k-1$ elements, as each pass on $Q$ pushes at most one pair. Additionally, as not all pairs are always added, the sizes are monotonically decreasing. On GPU, we exploit this observation to pre-allocate the correct queue size and guarantee that we will always have the necessary space.
To efficiently parallelize STQ on GPU, we exploit shared memory per thread block. In our experiments, the most efficient strategy consists of splitting the sorted boxes $B'$ into $m$ blocks containing 32 queries each (i.e., $m=k/32$). Since a warp consists of 32 threads, choosing a smaller number will introduce unnecessary branching. Using larger thread blocks can lead to more imbalanced workloads and longer runtime of branching in the code. The GPU thread-block scheduler performs well even with a larger grid size of thread blocks.

When running our algorithm on GPU, we need to prepare the data for the narrow-phase; this requires:
\begin{enumerate}
    \item splitting the set $C$ into edge-edge and vertex-face cases (SO),
    \item transforming the pairs in $C$ into narrow-phase data (CD),
    \item coping the vertex coordinates to the device (V2G).
\end{enumerate}
Similar to copying the boxes to GPU, these intermediate stages are negligible (\cref{fig:our-timings}).

\subsection{Narrow-Phase}\label{sec:m:np}
Since our algorithm builds upon \cite{Bolun2021Benchmark}, we first provide a self-contained overview of the original algorithm.

\paragraph{Summary of \cite{Bolun2021Benchmark}} The algorithm uses inclusion functions to detect collisions. It starts by constructing the interval $I = I_t\times I_{uv} \subseteq [0,1]^3$, where $I_t=[0, 1]$ is the time interval and $I_{uv}$ is the parameterization of the space. For a point-face query $I_{uv} = \{(u, v) \mid 0 \leq u, v \leq 1 \wedge u + v\leq 1\}$, where $u$ and $v$ are the triangle's barycentric coordinates. For an edge-edge query $I_{uv} = [0, 1]^2$, where $u$ and $v$ are the first and second segment parameter. Using $I$, the algorithm defines the box $B=B_F(I)$~\cite[Equation (4)]{Bolun2021Benchmark} that, if it intersects with a tolerance box $C_\epsilon$~\cite[Equation (5)]{Bolun2021Benchmark}\footnote{$C_\epsilon$ relies on floating-point operations to be compliant with the IEEE 754 standard, which is the case for GPU  (see \url{https://docs.nvidia.com/CUDA/floating-point/index.html}). Our algorithm is designed to account for the rounding error produced by these operations. Our floating-point filters are conservative in the sense that no rounding can make them fail. Possible \ac{FMA} contractions would only make the results of single operations more precise, and therefore our filters still work.
}, determines if the interval $I$ contains a root or not. In case it does, the algorithm recursively splits $I$ into two subintervals~\cite[Algorithm 2]{Bolun2021Benchmark} until they are either too small or completely contained in $C_\epsilon$.

\paragraph{Algorithm}

\begin{algorithm}
\begin{algorithmic}[1]
\Function{NarrowPhase}{$C, M_0, M_1$}
    \State $Q \gets $\Call{BuildIntervals}{$C, M_0, M_1$}\label{line:build_i}
    \State $t^\star \gets \infty$
    \While{$Q \neq \emptyset$}
        \State $Q' \gets \{ \}$
        \ForAll{$I\in Q$}\Comment{In parallel (CUDA kernel)}\label{line:process}
            \State $t^\star, I^l, I^r\gets $\Call{ProcessInterval}{$I, t^\star$}
            \If{$I^l \neq \emptyset$}
                \State $Q' \gets Q' \cup \{I^l\}$
            \EndIf
            \If{$I^r \neq \emptyset$}
                \State $Q' \gets Q' \cup \{I^r\}$
            \EndIf
        \EndFor
        \State $Q \gets Q'$\label{line:swap}
    \EndWhile
    \State \Return $t^\star$ \label{line:np:return}
\EndFunction
\State
\Function{ProcessInterval}{$I, t^\star$}\label{line:process-func}
    \State $t \gets I_t^l$
    \If{$t \geq t^\star$} \Comment{Current interval is after $t^\star$}\label{line:larger-than-toi}
        \State \Return $t^\star, \emptyset, \emptyset$
    \EndIf

    \State $B \gets B_F(I)$
    \If{$B \cap C_\epsilon = \emptyset$}\Comment{$I$ does not have collision}\label{line:skip}
        \State \Return $t^\star, \emptyset, \emptyset$
    \EndIf

    \If{$w(B) < \delta$ or $B \subseteq C_\epsilon$}\Comment{Collision found}\label{line:stop}
        \State \Return $t, \emptyset, \emptyset$ \label{line:np:collision-found}
    \EndIf

    \State $I^l, I^r \gets$ \Call{split}{I}\Comment{Refine $I$}
    \State \Return $t^\star, I^l, I^r$
\EndFunction

\end{algorithmic}
    \caption{Overview of the narrow-phase.}
    \label{alg:np}
\end{algorithm}

The method of \citet[Wang et al.]{Bolun2021Benchmark} can be trivially parallelized on a CPU by adding a parallel loop around the candidates $C$. This strategy works well on the CPU, but it is unfortunately not suitable for GPU architectures. For good performance, GPU architectures require all threads to perform the same operations and have similar memory access patterns. This is not the case for \cite{Bolun2021Benchmark}, as each query requires a different number and type of subdivisions.

To overcome this limitation, we observe that the core of the algorithm processes intervals and not queries (\cref{alg:np}). We can thus parallelize over interval splits instead of queries. This observation leads to performing the same operations on every thread independently from the candidate\footnote{An additional subtle benefit of this algorithm is that it makes the GPU kernel shorter, reducing the number of registers used, which is a common performance bottleneck on GPUs, where each streaming multiprocessor (SM) has a very small pool of registers available.}.

We start by constructing, for every collision candidate $C$, the initial interval $I=[0,1]^3$ (line~\ref{line:build_i}) and append them to the input queue $Q$. We then process in parallel all intervals in $Q$ and produce the output intervals' queue $Q'$ (line~\ref{line:process}). In the end, we swap the roles of the two queues (line~\ref{line:swap}) and continue alternating until $Q$ is empty.

\paragraph{Time of Impact}
We modified how we process a single interval (line~\ref{line:process-func}) with respect to~\cite{Bolun2021Benchmark} to account for the \ac{TOI}. We first check if $I_t^l$ (i.e., left-hand-side of the time interval of $I$) is larger than the current \ac{TOI} $t^\star$. In this case, $I$ can safely be skipped (line~\ref{line:larger-than-toi}).
Then we proceed as in~\cite{Bolun2021Benchmark} and, if the box $B$ constructed from $I$ does not intersect $C_\epsilon$, $I$ can be discarded as it does not contain a root (line~\ref{line:skip}).
Finally, if the width $w(I)$ of $I$ is smaller than a user-provided tolerance $\delta$ (or if $B$ is contained in $C_\epsilon$), we report a collision by returning $I_t^l$ (line~\ref{line:stop}).
If it is not the case, we split $I$ into a left $I^l$ and right $I^r$ interval and return the current time-of-impact as optimal.

\paragraph{Discussion}
The algorithm has only two necessary synchronizations: 1) the update of $t^\star$ \revv{using \texttt{atomicMin}}\footnote{\revv{While \texttt{atomicMin} does not directly support floating-point numbers we can use it by reinterpreting the bits in a double as a 64-bit signed integer using \texttt{\_\_double\_as\_longlong} as discussed here: \url{https://stackoverflow.com/a/51549250/13206140}.}} and 2) appending intervals to $Q'$. To efficiently append intervals, we keep track of the size of $Q'$ and use \texttt{atomicAdd} to increase the size counter when appending new elements.

\subsection{Batching}\label{sec:batching}

Running CCD on a GPU is particularly challenging as the amount of required memory might easily exceed the device's physical memory. For instance, the queries necessary to run the puffer ball scene cannot fit in \SI{12}{\giga\byte}. While splitting the work in \emph{batches} is possible for our method, it is challenging for other methods that use more complex spatial data structures. While other methods could have implemented batching, we are not aware of any existing CCD implementation that has this feature.

As our method checks every possible pair, we can schedule their execution in batches whose size depends on the available memory. To ensure that we have enough memory, we measure the size in bytes of the different data structures. Namely, let
\begin{itemize}
    \item $\S_P$ the size of the input parameters for the narrow phase (56 bytes);
    \item $\S_Q$ the size of the narrow phase queries (24 doubles per query);
    \item $\S_I$ the size of the narrow phase interval (252 bytes);
    \item $\S_i$ the size of two integers used to store a colliding pair (8 bytes).
\end{itemize}

We assume we can fit all boxes $b_i$ and the scene in CPU memory. Our batching strategy requires storing all boxes (but not collision pairs) on the GPU: this is not an issue in our experiments as we can fit  97,612,893 boxes on \SI{4}{\giga\byte} of memory, an unlikely scenario even for large scenes. In case the scene is larger, our code falls back to a CPU implementation. 
Once we construct the boxes, we estimate the available memory $\M$ using a CUDA function.
To ensure that we can run our algorithm safely, we construct the output queue $C$ containing the colliding pairs of maximum size $\S_C = (\M - \S_P)/ (\S_Q + \S_i)$. To compute the maximum size $\S_C$, we subtract the necessary memory for storing parameters $\S_P$; then we divide by the memory necessary to run every query in the narrow phase. Every query requires $\S_i$ bytes to store the results of the broad phase
and $\S_Q$ bytes to store the input query for the narrow phase. This ensures that if the broad phase appends at most $\S_C$ results, we can \emph{always} run the narrow phase (even if the queue $Q$ might overflow).

While running our broad phase, we append the colliding pair to the $C$ only if it does not overflow (in this case, we can run the whole algorithm without batching). If it does, we stop the broad phase and divide the input boxes into two batches and restart the algorithm until all batches can append all pairs to $C$.

In the narrow phase, the only problem that might occur is that the working queue $Q$ might overflow as we split the input intervals. We allocate $Q$ of size $\M / \S_I$, where $\M$ is the available memory after running the broad phase.
If $Q$ overflows, we discard the results and re-run the narrow phase using half of the pairs $C$. We repeat this procedure until $Q$ does not overflow. We note that we always keep the most accurate \ac{TOI} when we batch

\subsection{Guarantees}
\label{sec:guarantees}
Our broad-phase and narrow-phase algorithms are both \emph{conservative}. \Cref{sec:results-toi} empirically confirms that the final \ac{TOI} is also \emph{conservative} (i.e., less or equal to the exact value) while providing a measure of its difference from the ground truth.

\paragraph{Broad-phase} In \cref{alg:bp}, any intersection check amounts to only comparing floating point numbers. No other operation and/or rounding is involved in the process, meaning that intersection checks are all exact and $C$ contains all and only the intersecting pairs.
However, the boxes themselves might be slightly larger than necessary due to the rounding from double to single precision. Thus, our broad phase detection is \emph{conservative} because $C$ might include intersecting pairs that would not intersect without rounding. However, since the distortion is as small as machine precision, these spurious pairs are too few (if any) to have any perceivable impact on the overall performances.

\paragraph{Narrow-phase} As explained in \cref{alg:np}, if the two meshes collide, our process returns a finite \ac{TOI} (line~\ref{line:np:return}). This occurs if at least one root was found during the interval processing (line~\ref{line:np:collision-found}). In turn, this occurs if the condition on line~\ref{line:stop} is verified. This condition is the same as used by \citet[Wang et al.]{Bolun2021Benchmark}.
Our algorithm differs from \citet[Wang et al.]{Bolun2021Benchmark} because some additional intervals are discarded, whereas some others are added.
However, we discard an additional interval only if the current \ac{TOI} is finite (line~\ref{line:larger-than-toi}), meaning that a collision was already detected. Stated differently, our additional discards cannot determine a false negative.
Similarly, \citet[Wang et al.]{Bolun2021Benchmark} do not process an interval that we process only if the algorithm exits before having processed all the pairs and their intervals, which happens only if a collision is detected. That is, our additional intervals cannot determine a collision unless \citet[Wang et al.]{Bolun2021Benchmark} do the same. In essence, our algorithm returns a finite \ac{TOI} if and only if \citet[Wang et al.'s]{Bolun2021Benchmark} algorithm reports a collision and, since \citet[Wang et al.'s]{Bolun2021Benchmark} algorithm is conservative, we can conclude that ours is equivalently conservative.

\section{Benchmark}\label{sec:comparison}

We run all our experiments on an AMD Ryzen™ Threadripper™ PRO 3995WX 64-Cores @ \SI{2.7}{\giga\hertz} processor with 64 threads, \SI{512}{\giga\byte} of RAM, and an NVIDIA\textsuperscript{®} 3080 Ti. Note that, to avoid duplicated plots, we include the results of our method; we refer to \cref{sec:method} for a detailed explanation of what it does and to \cref{sec:results} for a discussion. The BVH method provides a function to update the data structure with new positions instead of rebuilding it from scratch at every frame, which we use in our experiments. For all other methods, the acceleration data structure is rebuilt at each frame.

\subsection{Broad-Phase}
\begin{figure*}
    \scriptsize\centering
    \parbox{0.01\linewidth}{\hfill}
    \parbox{.17\linewidth}{\centering Armadillo-Rollers}\hfill
    \parbox{.16\linewidth}{\centering Cloth-Ball}\hfill
    \parbox{.16\linewidth}{\centering Cloth-Funnel}\hfill
    \parbox{.16\linewidth}{\centering N-Bodies}\hfill
    \parbox{.16\linewidth}{\centering Rod-Twist}\hfill
    \parbox{.16\linewidth}{\centering Puffer Ball}\\[0.33em]
    \includegraphics[width=\linewidth]{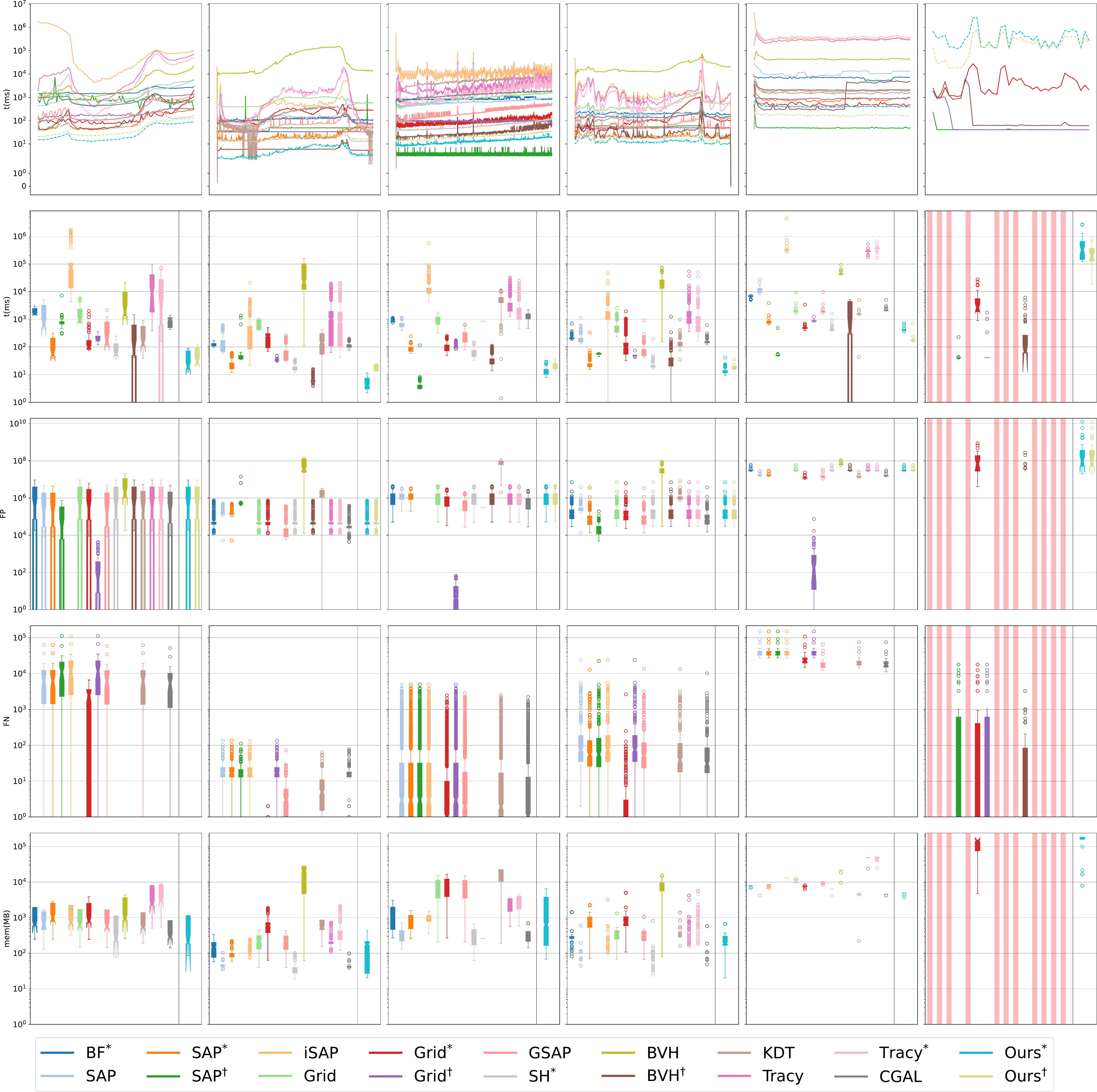}
    \caption{Results of the {\numOtherMethods} different methods for our six different scenes (columns). For each scene (\cref{fig:dataset}), we report the performance for every simulation frame (first row), timings box plot (second row), false positive box plot (third row), false negative box plot (fourth row), and memory box plot (fifth row). The star$^*$ depicts parallel CPU methods, while we use the dagger$^\dagger$ for GPU methods. For many methods, the puffer ball experiment run out of memory; we discarded them and marked them in red. The box plot shows aggregated statistics: the box extends from the first to the last quartile, the line in the middle is the median, and the lines (whisker) extend to the largest/smallest non-outlier point. Outliers are plotted as circles and defined as data points laying outside the 1.5 times interquartile range.}
    \label{fig:stats}
\end{figure*}

\Cref{fig:stats} presents detailed statistics of twelve implementations of a broad-phase collision detection algorithm run on our benchmark scenes.
For each broad-phase method, we compare the list of candidates with the ground truth data. We report the number of false positives (collisions detected by the method but are not colliding) and false negatives (collisions not detected by the method but are colliding), as well as the running time and maximal memory usage.
The number of false positives is quite stable among the methods, with a lower number for the Grid-based and the BVH-based approaches. All methods use a similar amount of memory, except for the BVH (for small scenes) and Tracy (for medium scenes).
We note that for SH, memory (and runtime) vary heavily based on the choice of the grid size (\cref{sec:sh-voxel}). The runtime varies significantly across the different methods, but the parallel SAP, SH, and GPU BVH are faster.

Since the goal of a broad-phase CCD is to filter unnecessary collision pairs, it is expected to introduce false positives, but it should be conservative; that is, it should \emph{never} have false negatives. In our experiments, despite inflating the bounding boxes by 1\% to ameliorate numerical problems, we found that the public implementations we use of GSAP, GpuGrid, GpuSAP, KDT, SAP, and CGAL have false negatives\footnote{Note that CGAL \texttt{box\_self\_intersection\_d} is fast but has a bug. The \texttt{box\_self\_intersection\_all\_pairs\_d} method is correct, but its quadratic complexity makes it intractable.}.
The false negatives produced by the GpuGrid and GpuSAP methods are at least in part due to the implementation choice of allocating a fixed-size array for candidates and ignoring any additional candidates. We believe that it is the case, as these methods are designed for rigid bodies where the number of boxes is proportional to the number of objects in the scene (typically not above tens of thousands), while for deformable bodies, the number of boxes depends on the complexity of the surface (i.e., one box per triangle, edge, and vertex) leading to a much larger number of boxes (the smallest scene in our dataset has 50 thousand boxes). This problem could be fixed by a more complex and less efficient implementation.

\subsection{Narrow-Phase}\label{sec:np-comparison}
For the narrow-phase, we compare only with the original CPU TI~\cite{Bolun2021Benchmark} and the \ac{ACCD} method of \citet[Li et al.]{Li2021CIPC} as it has been shown that \revv{all other algorithms prior to~\cite{Bolun2021Benchmark}} are either not conservative or produce a large number of false positives~\cite[Table 1]{Bolun2021Benchmark}.

\revv{A recent alternative, \citet[Lan et al.]{Lan2022Penetration} propose the \ac{FRA} \ac{CCD} and claims it is ``up to 60\% faster than [\ac{ACCD}].'' However, they do not release an open-source implementation. We attempt to implement the method ourselves\footnote{\url{https://github.com/zfergus/fast-approximate-root-ccd}}, but we quickly find many false negatives. While ours is a faithful implementation of the paper's algorithmic description, there may be technical implementation details missing from the paper that are required for the algorithm to work correctly. However, without the original authors' implementation, it is difficult to know. Therefore, for fairness, we opt to not provide any further comparisons.}

\revv{We initially used the open-source implementation of \ac{ACCD} provided by \citet[Li et al.]{Li2021CIPC} and found for certain scenes (cloth-funnel and n-bodies) it produced false negative. However, upon further examination, we observe this only occurs when the configuration at $t=0$ contains intersections. This is a violation of the assumptions made by \citet[Li et al.]{Li2021CIPC} who assume (because of their use of \ac{IPC}) that the meshes will always be intersection-free. Therefore, for a fair comparison, we adjust their algorithm to return true if the initial distance equals zero\footnote{\revv{We release our modified version as part of the IPC Toolkit~\cite{ipc_toolkit}.}}. This eliminates all false negatives at the default conservative scaling factor of $s=0.1$.}

\revv{From a theoretical standpoint, \ac{ACCD} can miss collision as rounding errors occur in the computation of distances, but their use of large numerical tolerances (i.e., $s=0.1$) avoids false negatives in exchange for a larger number of false positives and error in the \ac{TOI}. One can compute a more accurate \ac{TOI} by using a smaller $s$, but
this can lead to false negatives.
In comparison, \ac{TI} and our method account for the rounding error in the computation which allows us to reduce the root finding tolerance $\delta$ without introducing false negatives~\cite[Figure 8]{Bolun2021Benchmark}} \ZF{A full analysis here may require to perform the same analysis as in \cref{fig:toi-error} for \ac{ACCD}.}

\Cref{tab:narrow} summarizes the results of our benchmark of narrow-phase algorithms. Note, for each narrow-phase algorithm, we only record the \ac{TOI} for the whole timestep.

\begin{table}
    \caption{Average narrow-phase runtime (\unit{\ms}), the total number of \acf{FP} (in thousands), and the total number of \acf{FN} for \ac{ACCD}, \ac{TI}, and Ours on GPU.\todo{Adjust the ACCD results here.}}
    \label{tab:narrow}
    \centering\footnotesize
    \begin{tabular}{ll|c|c|c}
\toprule
\thead{Scene} & \thead{Method} & \thead{Runtime (\unit{\ms})} & \thead{FP} & \thead{FN} \\
\midrule
\multirow{3}{*}{Armadillo-Rollers} & ACCD ($s=0.1$) &    2 &    8594 &       0 \\
          & TI  &   32 &    8594 &       0 \\
          & Ours  &    7 &    8594 &       0 \\
\midrule
\multirow{3}{*}{Cloth-Ball} & ACCD ($s=0.1$)  &    7 &   18637 &       0 \\
           & TI  &  103 &   18637 &       0 \\
           & Ours&   15 &   18637 &       0 \\
\midrule
\multirow{3}{*}{Cloth-Funnel} & ACCD ($s=0.1$)  &    1 &    2995 &    0 \\
             & TI  &   12 &    3417 &       0 \\
             & Ours  &    5 &    3417 &       0 \\
\midrule
\multirow{3}{*}{N-Bodies} & ACCD ($s=0.1$)  &  150 &  232991 &  0  \\
         & TI  &  402 &  232965 &       0 \\
         & Ours &   46 &  232965 &       0 \\
\midrule
\multirow{3}{*}{Rod-Twist} & ACCD ($s=0.1$)  &    4 &  457479 &       0 \\
          & TI  &  195 &  457479 &       0 \\
          & Ours &   24 &  457479 &       0 \\
\bottomrule
\end{tabular}
\end{table}

\section{Evaluation}\label{sec:results}

\revv{We implement our method in C++ and CUDA. We utilize Eigen~\cite{eigen} on CPU for linear algebra computations and oneAPI Threading Building Blocks (oneTBB)~\cite{TBB} to add parallelism to our CPU implementation.}

\subsection{Comparison}

\Cref{sec:comparison} provides an in-depth comparison of our method and the \numOtherMethods{} prior works.

Overall, our GPU algorithm is up to $20\times$ faster for larger scenes than the best existing combination (BVH for broad-phase and CPU parallel TI): $3\times$ faster for armadillo-rollers, $10\times$ for cloth-ball, $1.6\times$ for cloth-funnel, $22\times$ for n-bodies, and $17\times$ for rod-twist.

\paragraph{Broad Phase}
As for the other BP methods (\cref{fig:stats}), our method has similar accuracy to any other algorithm. The performance of our broad-phase methods is mostly independent of the scene: ours is consistently among the fastest methods. Our method on GPU is faster than BVH on large scenes (n-bodies and rod-twist) and has a comparable time for smaller ones. On CPU, our method has a similar performance to the spatial hash; however, it does not require tweaking the cell size (\cref{sec:sh-voxel}). Our method uses slightly more memory than BF, as it does not need to store any data structure.

\paragraph{Narrow Phase}
ACCD is $3.5\times$ faster than our method, but it fails to detect collisions, making it not suitable for contact methods using interior point optimization (e.g., IPC). TI is consistently slower than our method (between 2.4 and 8.7 times faster).
 \todo{Revise this section.}

\subsection{Scaling}

\begin{figure}
    \centering
    \includegraphics[width=.6\linewidth]{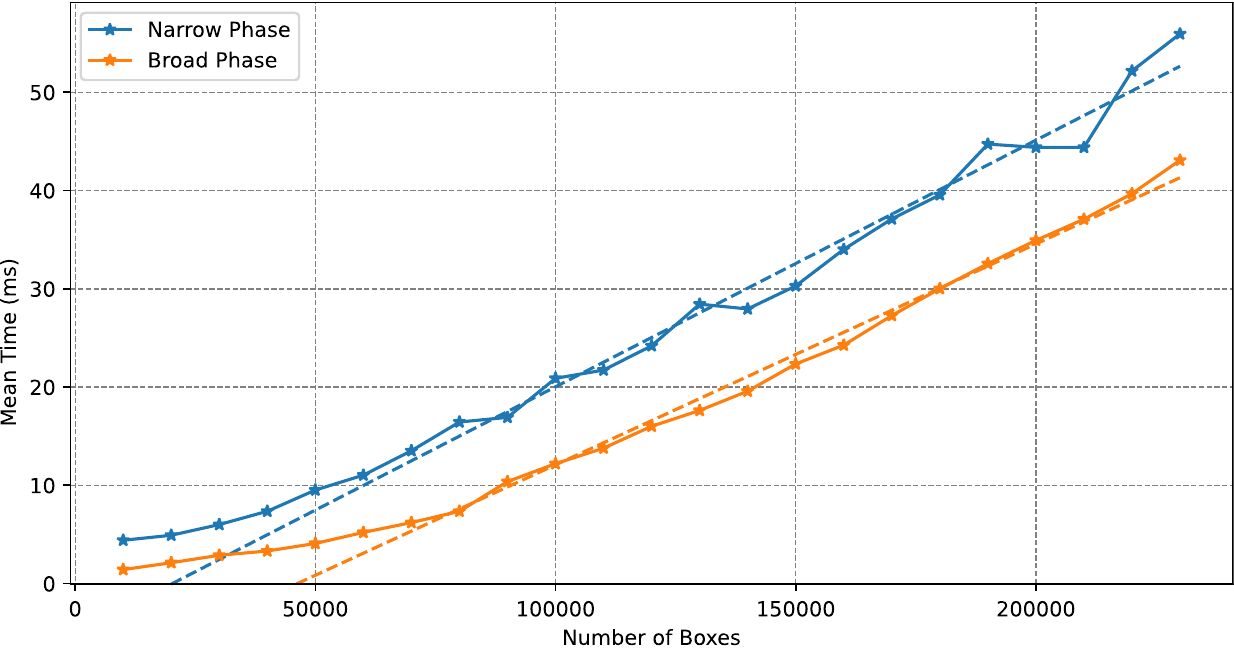}
    \caption{Run time of our algorithm with respect to the number of queries. To generate a varying number of queries, we select a random sub-sample of all possible boxes from the last 10 frames of rod-twist. }
    \label{fig:complexity}
\end{figure}

While our algorithm is inspired by a brute force approach which has quadratic complexity with respect to the number of boxes, we borrow ideas from the SAP to limit the average complexity of our algorithm. \Cref{fig:complexity} shows that the run time of both of our algorithms grows \revv{nearly} linearly with the number of boxes.
\revv{Our theoretical complexity is the same as SAP: $O(|B|\log(|B|)) + O(|B| + |C|)$ where $|B|$ is the number of boxes and $|C|$ is the number of overlaps~\cite{Baraff:1992:Dynamic}.}

\begin{figure}
    \centering
    \includegraphics[width=.6\linewidth]{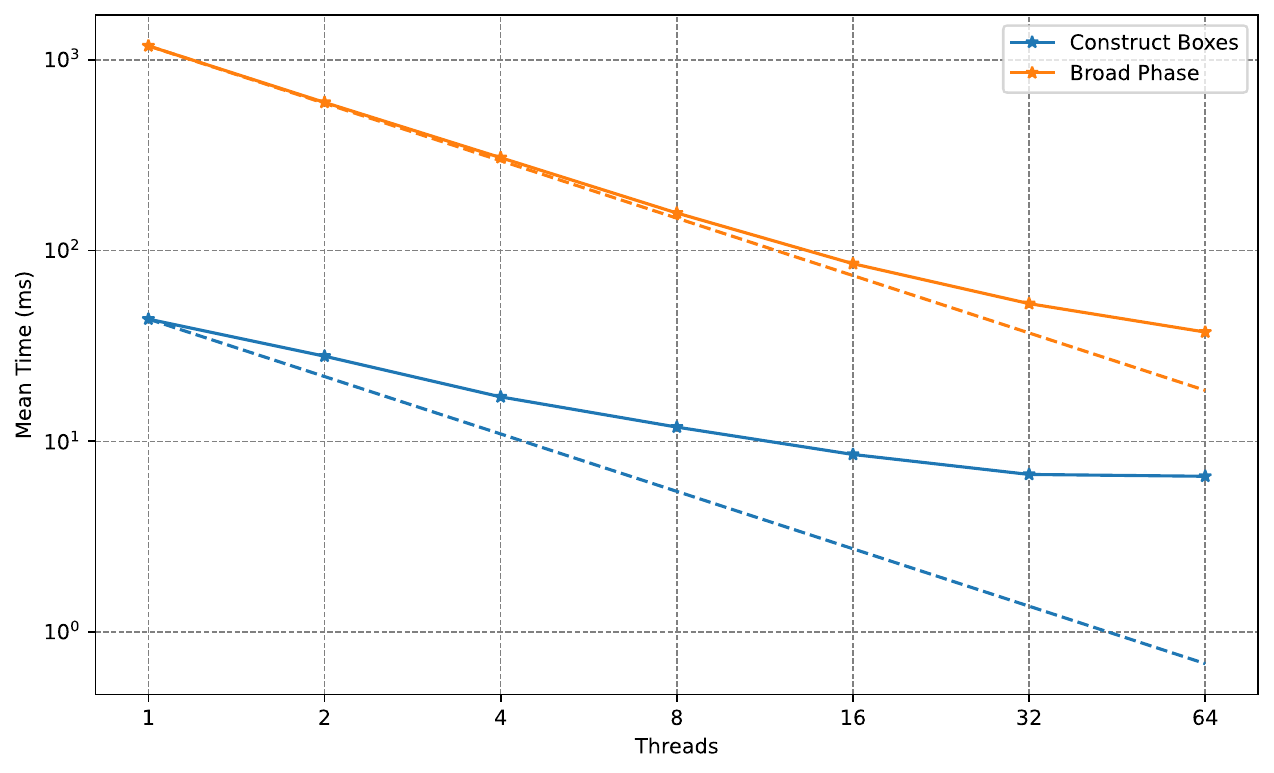}
    \caption{Strong scaling of our method for the last 10 frames of rod-twist. The dashed lines show the perfect scaling.}
    \label{fig:scalability}
\end{figure}

To assess the parallel scalability of our broad-phase method, we run the last ten frames of rod-twist on CPU, varying the number of threads from 1 to 32 (\cref{fig:scalability}). Our algorithm scales well with respect to the number of threads: with 8 threads, the broad-phase is 7.5 times faster, and it gets 22 times faster with 32 threads. Constructing the boxes scales the worst: 3.7 times faster with 8 threads and 6.5 faster for 32 threads. However, it is around one order of magnitude faster than the broad-phase, making it negligible.

\subsection{Time of Impact Validation and Accuracy}\label{sec:results-toi}
\begin{figure}
    \centering
    \includegraphics[width=\linewidth]{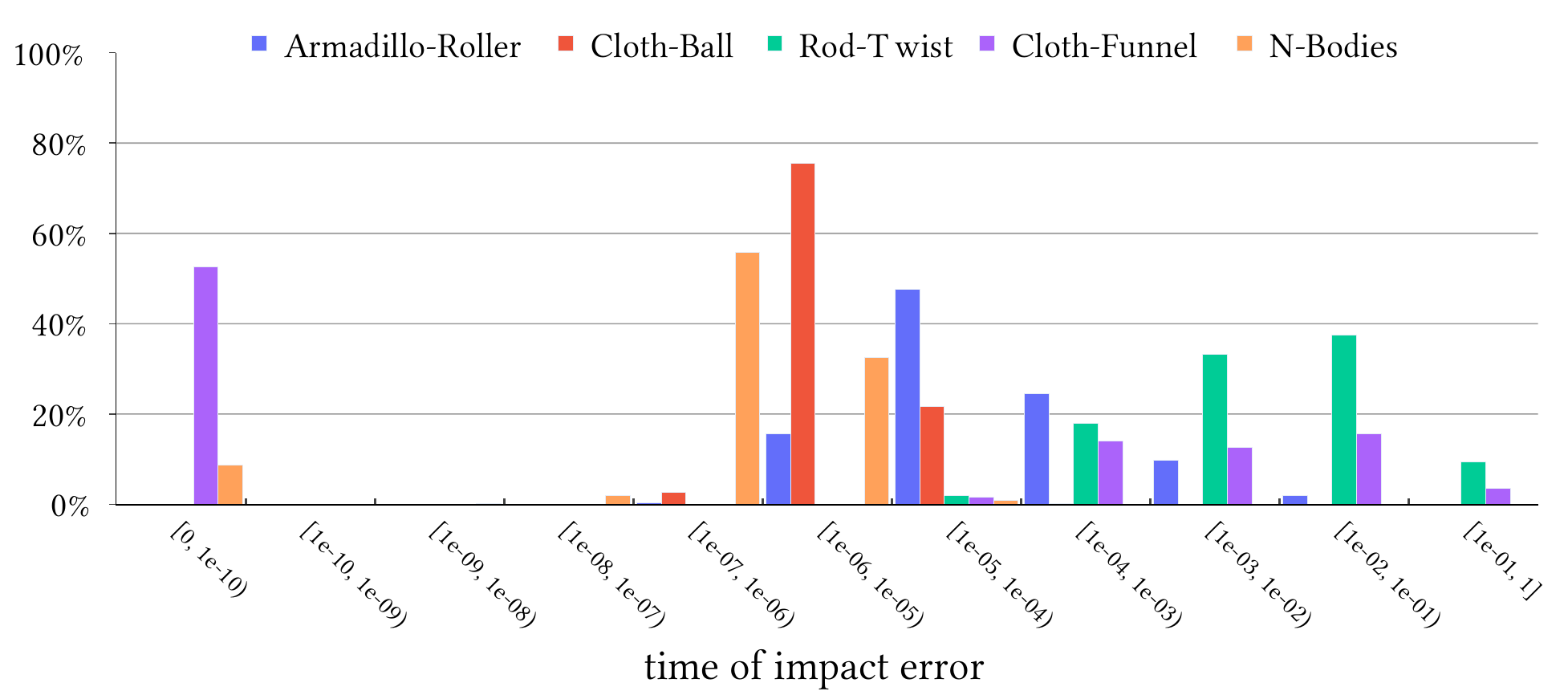}
    \caption{Using the symbolic ground truth for the \ac{TOI}, we compute an error per query per scene and plot them as a histogram in log scale. The error is computed as the absolute difference between the earliest root reported by the symbolic root finder and our code (the error is approximately computed using 128 bits of precision). All \acp{TOI} computed by our method are verified symbolically to be less than the ground truth.}
    \label{fig:toi-error}
\end{figure}

Taking a step size greater than the exact \ac{TOI} will lead to intersections, so we confirm our predicted \ac{TOI} is less than or equal to the symbolic expression (\cref{sec:dataset-toi}). This confirms our method is conservative. Additionally (and not used to verify correctness), we estimate the error of our \ac{TOI} by evaluating the difference using 128 bits of precision (\cref{fig:toi-error}).

The mean error for all queries is $0.0023$ with a standard deviation of $0.018$ and a median error of $\num{1.03e-6}$. We note that this distribution varies between scenes (e.g., cloth-ball has a mean error of $\num{8.33e-6}$ compared to $0.032$ for rod-twist). This indicates a dependency between the types of contact and the accuracy of the \ac{TOI}.

\subsection{Different Architectures}\label{sec:architectures}

\begin{figure*}
    \footnotesize\centering
    \includegraphics[width=\linewidth]{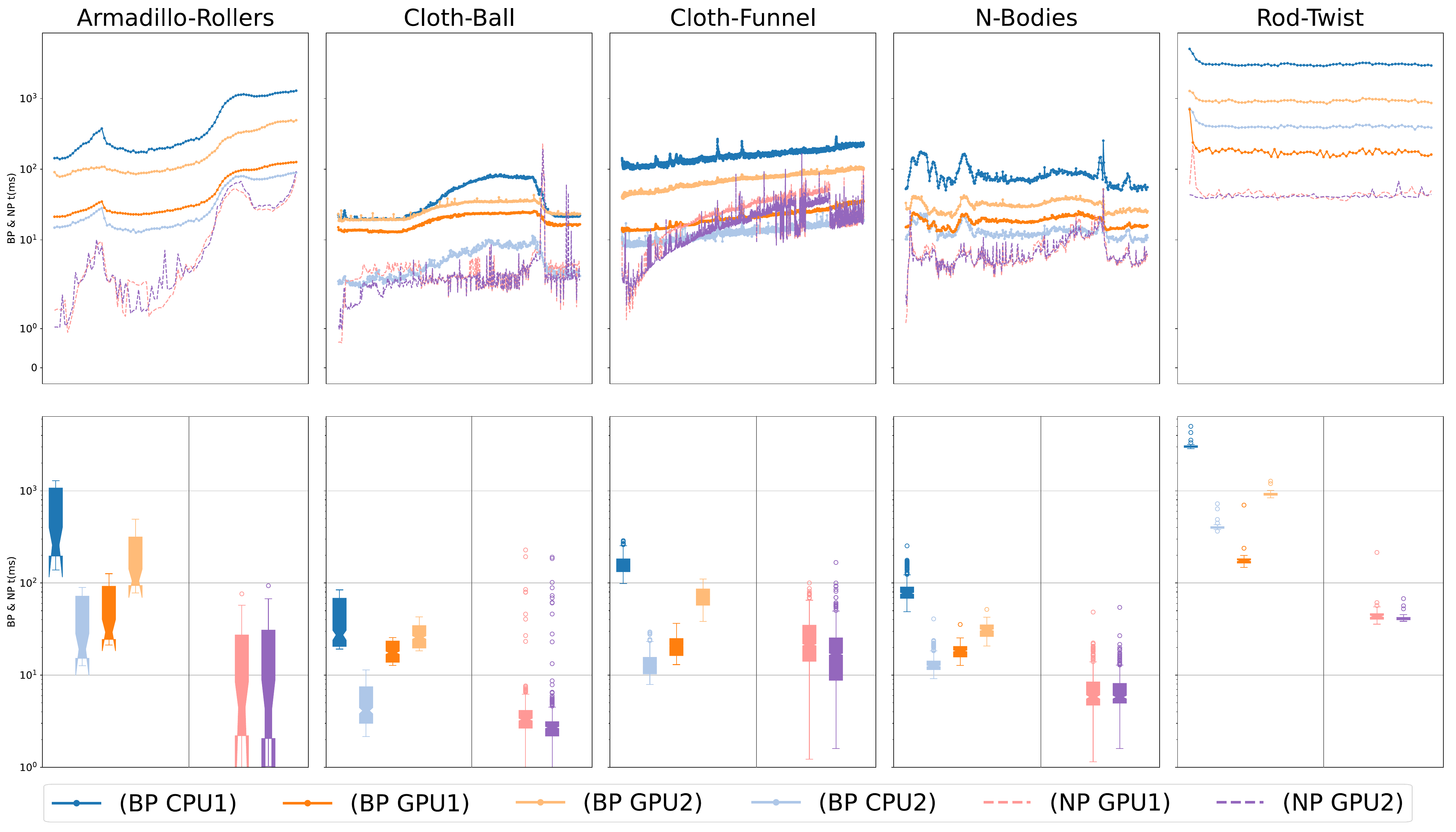}
    \caption{
    Results of our method on the consumer architectures (CPU1 and GPU1) and the professional architectures (CPU2 and GPU2) for five different scenes (columns). For each scene (Figure~\ref{fig:dataset}), we report the performance for every simulation frame (first row) and timings box plot (second row).
    }
    \label{fig:all-timings}
\end{figure*}

We also run our method on different ``workstation'' hardware (\cref{fig:all-timings}); two CPUs: CPU1 a consumer architecture (Intel\textsuperscript{®} Core™ i7-5930K CPU @ \SI{3.5}{\giga\hertz}) and CPU2 a professional CPU (AMD Ryzen™ Threadripper™ PRO 3995WX 64-Cores @ \SI{2.7}{\giga\hertz}) and two \acp{GPU}: GPU1 a consumer-grade card (NVIDIA\textsuperscript{®} 3080 Ti) and GPU2 a professional-grade card  (NVIDIA\textsuperscript{®} v100). Using this naming convention, the results in \cref{sec:comparison} are run on CPU2 and GPU1. For every CPU run, we limit the number of threads to 12. We note that for many computations, a CPU implementation on CPU2 has an execution time comparable to a GPU implementation\footnote{\revv{E.g., on the Blender Open Data benchmark (\url{https://opendata.blender.org/}), CPU1 has a median score of 85.02, CPU2 has a median score of 1067.03, GPU1 has a median score of 5421.29, and GPU2 has a median score of 1800.77. Higher values are better.}}: the hardware is, however, expensive as CPU2 costs around ten times more than a top-of-the-line GPU. 

The narrow phase on GPU is faster than the broad phase (just barely for the rod-twist and 9 times for the n-bodies); on CPU, the two phases are more comparable (narrow-phase is 3 times slower on rod-twist and 3 times faster on cloth-funnel). The difference comes from the fact that the narrow-phase on GPU is up to 80 times faster than CPU, while the broad-phase peaks at 10 times faster. As expected for our method on CPU2 has a similar performance on GPU1.

\begin{figure}
    \centering
    \includegraphics[width=\linewidth]{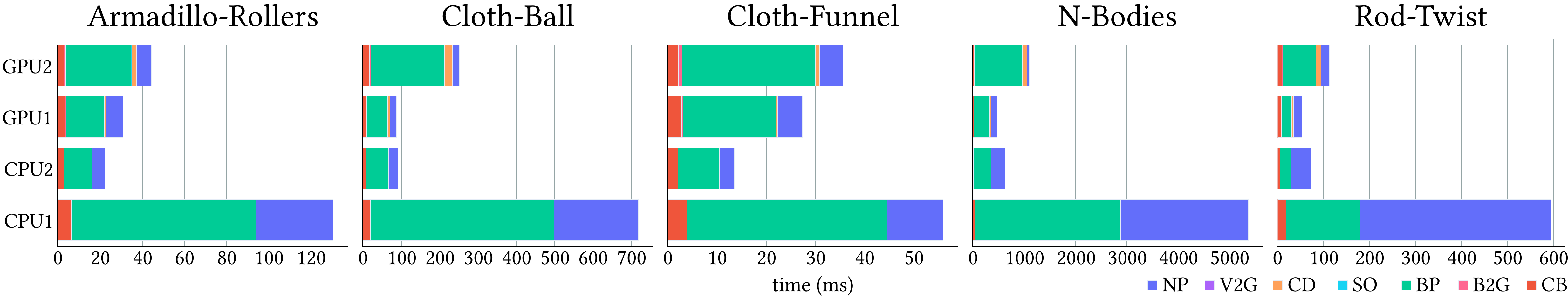}
    \caption{Breakdown of the runtime for every scene for our method on the different architectures. The parts are: narrow phase (NP), copy the vertices to the GPU (V2G), split the queries into vertex-face and edge-edge (SO), broad phase (BP), construct the data list (CD), copy the boxes to the GPU (B2G), and constructs the boxes (CB). \Cref{sec:architectures} details the different architectures.
    }
    \label{fig:our-timings}
\end{figure}

\Cref{fig:our-timings} shows the cutoff of the different phases of our algorithm; the broad phase dominates the computation, in particular on the slower CPU. For scenes with complex contacts (rod-twist and n-bodies) the narrow phase becomes more prominent. When switching to a faster architecture (CPU1 to CPU2 and \revv{GPU2 to GPU1}), both phases obtain a similar speedup. The narrow phase benefits more when switching from CPU to GPU. The other parts of our algorithm (e.g., memory copy, allocation, etc.) are negligible.

\subsection{Batching}
\begin{figure}
    \centering
    \parbox{0.8\linewidth}{\includegraphics[width=\linewidth]{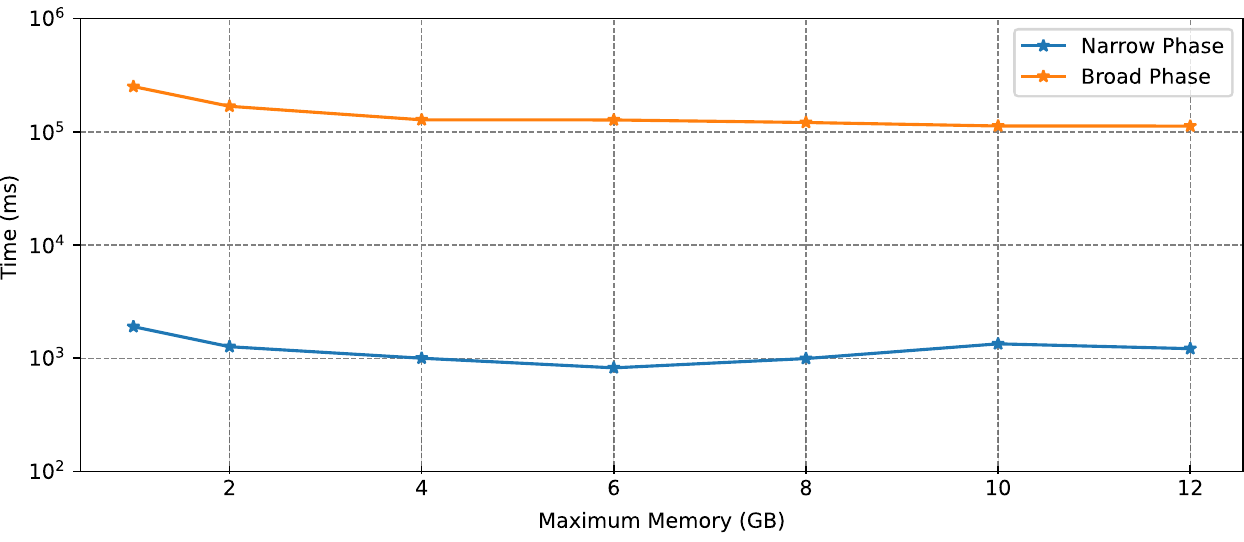}}\\
    \caption{Runtime of our algorithm for a scene with 13 billion queries (top) for different amounts of memory (bottom). 
    }
    \label{fig:batching}
\end{figure}

To evaluate the runtime overhead of our batching strategy (\cref{sec:batching}), we artificially limited the available memory on the GPU between 1GB and 12GB (\cref{fig:batching}). As we decrease the memory, our algorithm becomes slightly slower. For the narrow phase at 6GB, one of the early batches finds a small \ac{TOI}, leading to all other chunks to quickly terminate.

\section{Application to Simulation}\label{sec:sim}

\begin{table}[t]
\caption{Performance of our new CCD and broad phase for the unit tests of \citet[Erleben]{erleben2018methodology} in~\cite[Figure 11]{Li2020IPC}, the five cube stack, mat-twist, and mat-knives simulations. For tiny scenes the overhead of our method worsens performance, but in general leads to a performance increase in both the CCD and simulation as a whole. For larger scenes, the bottleneck shifts to the linear solves and Hessian matrix assembly leading to a smaller overall improvement of running time.}
\label{tab:erleben}
\setlength{\tabcolsep}{2pt}
\footnotesize
\centering
\begin{tabular}{@{}l|c|c|c|c|c|c|c|c@{}}
\toprule
\multirow{2}{*}{\thead{Scene}} & \multirow{2}{*}{\thead{\#V}} & \multirow{2}{*}{\thead{\#T}} & \multicolumn{2}{c|}{\thead{CCD Time (\unit{\s})}} & \multicolumn{2}{c|}{\thead{Total Time (\unit{\s})}} & \multirow{2}{*}{\thead{CCD\\Speed-Up}} & \multirow{2}{*}{\thead{Total\\Speed-Up}} \\
 & & & \thead{SH+TI} & \thead{Ours} & \thead{SH+TI} & \thead{Ours} &  & \\
\midrule
Spikes & 5 & 2 & 0.01 & 0.73 & 0.12 & 0.90 & \cellcolor[HTML]{FFC7CE}{\color[HTML]{9C0006} 0.01$\times$} & \cellcolor[HTML]{FFC7CE}{\color[HTML]{9C0006} 0.13$\times$}\\
Spike and Wedge & 5 & 2 & 0.01 & 0.79 & 0.13 & 0.89 & \cellcolor[HTML]{FFC7CE}{\color[HTML]{9C0006} 0.02$\times$} & \cellcolor[HTML]{FFC7CE}{\color[HTML]{9C0006} 0.14$\times$}\\
Spike in a Hole & 5 & 2 & 2.47 & 4.65 & 2.93 & 5.29 &\cellcolor[HTML]{FFC7CE}{\color[HTML]{9C0006} 0.53$\times$} & \cellcolor[HTML]{FFC7CE}{\color[HTML]{9C0006} 0.55$\times$}\\
Wedge in a Crack & 6 & 3 & 5.11 & 6.17 & 5.75 & 7.17 & \cellcolor[HTML]{FFC7CE}{\color[HTML]{9C0006} 0.83$\times$} & \cellcolor[HTML]{FFC7CE}{\color[HTML]{9C0006} 0.80$\times$}\\
Sliding Spike & 5 & 2 & 0.82 & 0.62 & 0.86 & 0.73 & \cellcolor[HTML]{C6EFCE}{\color[HTML]{006100} 1.31$\times$} & \cellcolor[HTML]{C6EFCE}{\color[HTML]{006100} 1.18$\times$}\\
Spike in a Crack & 5 & 2 & 5.45 & 3.58 & 5.80 & 4.13 & \cellcolor[HTML]{C6EFCE}{\color[HTML]{006100} 1.52$\times$} & \cellcolor[HTML]{C6EFCE}{\color[HTML]{006100} 1.40$\times$}\\
Cliff Edges & 8 & 6 & 3.98 & 1.66 & 4.46 & 2.24 & \cellcolor[HTML]{C6EFCE}{\color[HTML]{006100} 2.40$\times$} & \cellcolor[HTML]{C6EFCE}{\color[HTML]{006100} 1.99$\times$}\\
Internal Edges & 8 & 6 & 6.31 & 2.57 & 6.93 & 3.38 & \cellcolor[HTML]{C6EFCE}{\color[HTML]{006100} 2.45$\times$} & \cellcolor[HTML]{C6EFCE}{\color[HTML]{006100} 2.05$\times$}\\
Sliding Wedge & 6 & 3 & 1.78 & 0.50 & 1.85 & 0.58 & \cellcolor[HTML]{C6EFCE}{\color[HTML]{006100} 3.59$\times$} & \cellcolor[HTML]{C6EFCE}{\color[HTML]{006100} 3.21$\times$}\\
Wedges & 6 & 3 & 7.57 & 1.95 & 7.81 & 2.33 & \cellcolor[HTML]{C6EFCE}{\color[HTML]{006100} 3.88$\times$} & \cellcolor[HTML]{C6EFCE}{\color[HTML]{006100} 3.36$\times$}\\
5 Cubes (\cref{fig:5-cubes}) & 40 & 30 & 34.3 & 5.88 & 36.0 & 7.52 & \cellcolor[HTML]{C6EFCE}{\color[HTML]{006100} 5.83$\times$} & \cellcolor[HTML]{C6EFCE}{\color[HTML]{006100} 4.78$\times$} \\
Mat-Twist (\cref{fig:mat-twist}) & 3.2K & 9.1K & 7.92 & 2.60 & 2567.06 & 2368.87 & \cellcolor[HTML]{C6EFCE}{\color[HTML]{006100} 3.05$\times$} & \cellcolor[HTML]{C6EFCE}{\color[HTML]{006100} 1.08$\times$}\\
Mat-Knives (\cref{fig:codim}) & 3.2K & 9.1K & 86.1 & 25.2 & 178.1 & 146.7 & \cellcolor[HTML]{C6EFCE}{\color[HTML]{006100} 3.42$\times$} & \cellcolor[HTML]{C6EFCE}{\color[HTML]{006100} 1.21$\times$}\\
\bottomrule
\end{tabular}
\end{table}

We use our GPU CCD algorithm inside the IPC algorithm~\cite{Li2020IPC} implemented in PolyFEM~\cite{polyfem} by running several simulations on CPU2 with 8 threads for the simulation and GPU1 for the CCD (see \cref{sec:architectures} for a description of the architectures). We note that IPC requires computing the distances between primitives at the beginning of every time step, which we accelerate using our STQ broad phase algorithm. We run all the unit tests of \citet[Erleben]{erleben2018methodology} presented in~\cite[Figure 11]{Li2020IPC} using the original implementation (using a \ac{SH} for broad phase and the parallel \ac{TI} narrow phase CCD) and compare with our method (\cref{tab:erleben}).

\begin{figure}[t]
    \centering
    \includegraphics[width=0.8\linewidth]{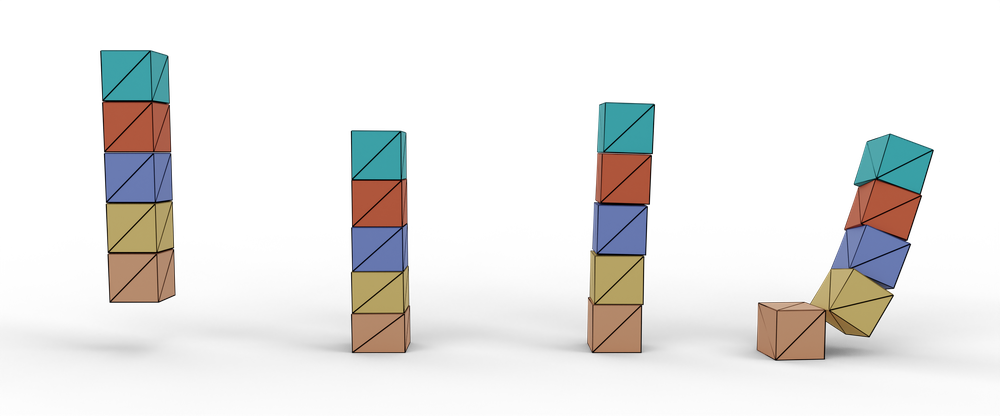}
    \caption{Several frames of the five-cube stack example. The whole scene has only 30 tetrahedra and is, therefore, CCD bound. We see a $5.83\times$ speed-up in CCD and $4.78\times$ overall compared to using SH and TI.}
    \label{fig:5-cubes}
\end{figure}

We also run the five-cube stack example (\cref{fig:5-cubes}) that contains several resting contacts. Similar to the unit tests, the meshes here are extremely coarse. When using our method, the simulation is $4.5\times$ faster. 

\begin{figure}
    \centering
    \includegraphics[width=0.8\linewidth]{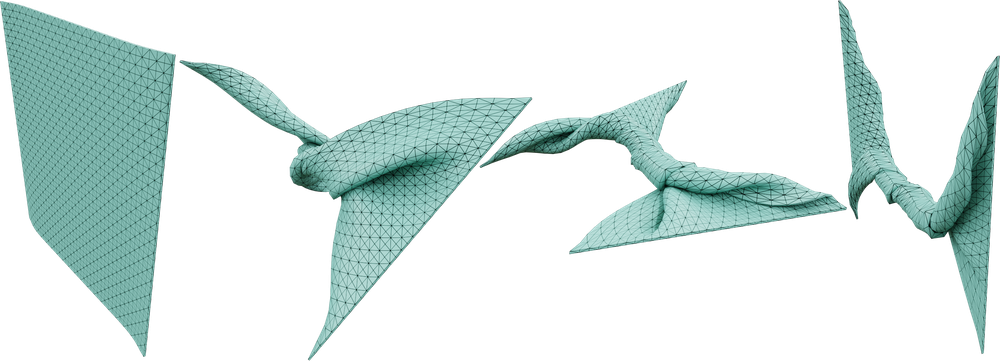}
    \caption{Several frames of a mat-twist simulation utilizing our scalable continuous collision detection (CCD) algorithm. We see a $3.05\times$ speed-up in the CCD and a $1.08\times$ speed-up overall compared to using SH and TI.}
    \label{fig:mat-twist}
\end{figure}

When using denser meshes (\cref{fig:mat-twist} has 9K tetrahedra) and the elastic deformations become more challenging, the non-linear elastic solver dominates the IPC runtime, and the speedup provided when using our method becomes less prominent (only 8\% times faster overall).

\begin{figure}
    \centering
    \includegraphics[width=0.8\linewidth]{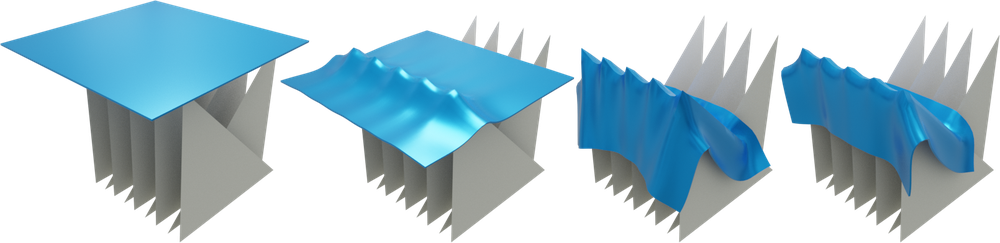}
    \caption{Several frames of a codimensional simulation. The mat has 9K tetrahedra, while the codimensional triangles are not deformable. We see a $3.42\times$ speed-up in the CCD and a $1.21\times$ speed-up overall compared to using SH and TI.}
    \label{fig:codim}
\end{figure}

To stress-test our CCD algorithm, we show our method is able to handle CCD between codimensional objects in \cref{fig:codim}. In this scene, we again see similar speedups as \cref{fig:mat-twist} with a 21\% speed-up overall.

\subsection{Avoiding Time of Impacts Equal to Zero}
An important caveat of using a conservative CCD method inside the IPC algorithm is that it should not produce a \ac{TOI} $t^\star$ of zero. A $t^\star$ of zero causes the non-linear solver to stagnate because IPC uses the $t^\star$ to determine the maximum step size allowable inside the optimization-based implicit time-stepping. Additionally, IPC guarantees every step results in an intersection-free state, so $t^\star$ cannot be zero (i.e., not initially intersecting).

An exact CCD method would provide this guarantee. However, our method is conservative, so even if the objects are not touching, a na\"ive implementation can produce a \ac{TOI} of zero.

To avoid $t^\star=0$, we make slight modifications to \cref{alg:overview,alg:np} (\cref{app:ipc-changes}). As part of the strategy to avoid $t^\star = 0$, the IPC algorithm uses a minimum separation in the CCD to prevent taking a step that results in parts exactly touching. We choose the minimum separation relative to the initial distance $d_0$ between the query's primitives. We use $0.2d_0$ in all our experiments. To implement minimum separation CCD, we use the same strategy as TI~\cite{Bolun2021Benchmark}: we enlarge the box $C_\epsilon$ by the minimum separation distance.

\section{Conclusion}

We introduce a novel dataset that provides a way to check for the correctness of CCD codes and their \ac{TOI} in different settings. We believe it is a realistic and practical approach to evaluate the conservativeness of CCD implementations, even if passing the benchmark is not a formal proof of correctness. It helped us design our algorithm; using it, we found counter-examples for other CCD codes. The benchmark is easy to extend, and we plan to keep it up to date and add more scenes and challenging queries in the following years.

Based on the benchmark, we designed a novel scalable CCD algorithm combining broad and narrow phase collision detection. Our algorithm is provably conservative, and our implementation has been tested on multiple combinations of recent operating systems and hardware architectures. Our algorithmic contribution specifically targets parallel architectures with high memory bandwidth (and high latency), which have very different requirements than traditional serial architectures. Our algorithm scales well to GPU hardware: an NVIDIA\textsuperscript{®} 3080 Ti GPU (MSRP ${\sim}1.2$K USD) achieves a speed comparable to a CPU server chip with 64 cores/128 threads (MSRP ${\sim}10$K USD). We believe that our GPU algorithm could be extended to run on multi-GPU. Our preliminary experiments show that the broad phase becomes 2.3 times faster when using 4 GPUs for the N-Bodies scene.

When integrated with the state-of-the-art solver IPC, our approach reduces the overall simulation time, which we believe is of practical relevance to the graphics and simulation community. Our implementation is released on GitHub under the MIT license to aid in its adoption in academia and industry.

\bibliographystyle{siamplain}
\bibliography{99-biblio}

\appendix
\section{Spatial hash voxel size}\label{sec:sh-voxel}

\begin{figure}
    \centering
    \includegraphics[width=\linewidth]{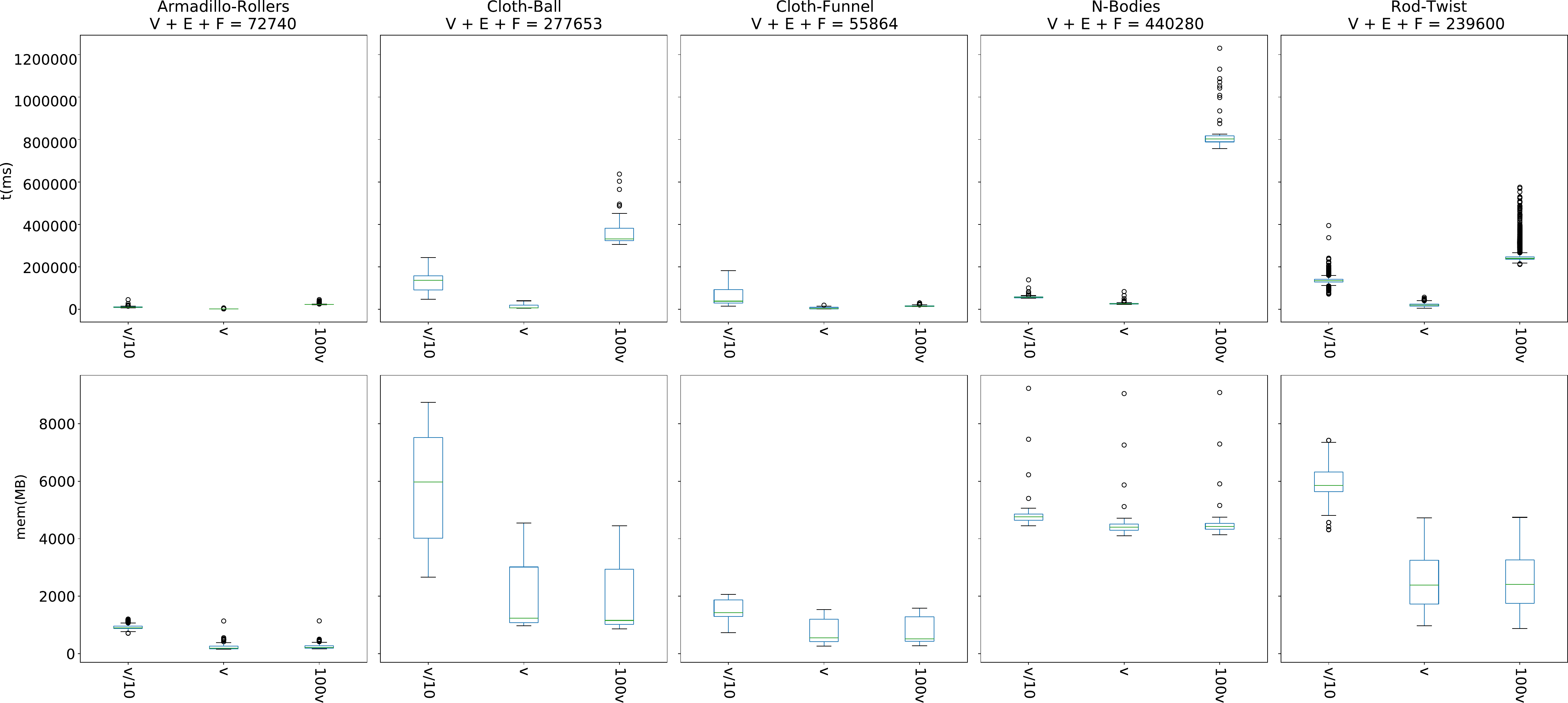}
    \caption{\textbf{SH Voxel Size.} We benchmark SH using three different values for the voxel size and plot the timing (top) and memory (bottom). We use a voxel size of $v/10$, $v$, and $100v$, where $v$ is the heuristic size used throughout all of our experiments. We also experiment with smaller voxel sizes but run out of memory for even $v / 20$ (max \SI{64}{\giga\byte}) due to many duplicate collision candidates.}
    \label{fig:sh-voxels}
\end{figure}

The performance and memory footprint of SH heavily depends on the voxel size (\cref{fig:sh-voxels}). We use $v=2 \max(d_0, d_1)$, with $d_0$ the average edge length and $d_1$ the average displacement as an empirical heuristic for our experiments. Using smaller or larger voxels leads to an increase in runtime (with large voxels being the worst); however, too small voxels consume too much memory and eventually run out of memory (for $v / 20$).

\section{Zero time of impact and minimum separation}\label{app:ipc-changes}

To avoid $t^\star = 0$, we make slight modifications to \cref{alg:overview,alg:np}. As in \cite{Li2020IPC}, if \cref{alg:overview} return $t^\star=0$ we perform the narrow-phase again but set the minimum separation to 0 and enable a no zero ToI strategy.

This no-zero ToI strategy dictates that if $I_t^l=0$, $I$ should always be split (ignoring user tolerances and the maximum number of splits). We note that under floating-point division this split might not be possible, but this has not been encountered in practice and would most likely require a degenerate case involving tiny distances (which IPC does a good job of preventing thanks to its barrier potential method of handling contacts). In the end, because the minimum separation was disabled, the resulting $t^\star$ is multiplied by a scaling factor less than 1 to avoid exactly touching after the step (we use 0.8 in our examples).

\end{document}